\documentclass[12pt]{article}
\pdfoutput=1
\usepackage{jheppub}
\usepackage[utf8]{inputenc}
\usepackage{verbatim}
\usepackage{amsmath}
\usepackage{amssymb}
\usepackage{amsthm}
\usepackage{amsfonts}
\usepackage{hyperref}
\usepackage{MnSymbol}
\usepackage{dsfont}
\usepackage{xcolor}
\newcommand{\be}{\begin{equation}}
\newcommand{\ee}{\end{equation}}
\newcommand{\ran}{\rangle}
\newcommand{\lan}{\langle}
\newcommand{\mO}{\mathcal{O}}

\newcommand{\wt}{\widetilde}
\newcommand{\bi}{\begin{itemize}}
\newcommand{\ei}{\end{itemize}}
\newcommand{\Tr}{\mathrm{Tr}}

\newcommand{\M}{\mathcal{M}}
\newcommand{\bfig}{\begin{figure}\begin{center}}
\newcommand{\efig}{\end{center}\end{figure}}

\newcommand{\T}{\mathcal{T}}
\newcommand{\I}{\mathrm{I}}

\usepackage{graphicx}
\usepackage{physics}
\usepackage[dvipsnames]{xcolor}

\DeclareMathOperator{\arctanh}{arctanh}

\begin{document}
\title{Anti-scrambling and euclidean folds from observer correlators in de Sitter space}
\author[a]{Daniel Harlow}
\author[a]{Ying Zhao}
\affiliation[a]{Center for Theoretical Physics - a Leinweber Institute\\ Massachusetts Institute of Technology, Cambridge, MA 02139, USA}
\emailAdd{harlow@mit.edu}
\emailAdd{zhaoying@mit.edu}
\abstract{The de Sitter horizon behaves in a qualitatively different way from a black hole horizon, which poses a challenge to any attempt to develop a fundamental quantum description of de Sitter space.  In this paper we gather some ``data'' on this problem using gravitational calculations, seeing that they lead to an ``anti-scrambling'' phenomenon that is contrary to the behavior of standard many-body quantum systems.  We organize our discussion in terms of correlation functions computed on the worldline of an observer living in the spacetime, with the two-point function looking like a conventional thermal correlator but the four-point function showing anti-scrambling.  We propose that anti-scrambling can be realized in a quantum system whose Hamiltonian is bounded from both above and below using correlators that are folded in Euclidean time.}

\maketitle
\section{Introduction}
One of the main insights about the quantum mechanics of black holes is that, seen from the outside, they behave as fairly conventional many-body quantum systems.  More concretely, we can probe a black hole by perturbing it and seeing how it responds; the response can be organized in terms of a set of correlation functions that resemble those of a many-body quantum system.  In particular, this includes thermalization effects that are probed by two-point functions \cite{Maldacena:2001kr,Son:2002sd} and scrambling effects that are probed by four-point functions \cite{Shenker:2013pqa,Roberts:2014ifa,Shenker:2014cwa}.  In the context of negative cosmological constant this agreement between black holes and many-body quantum mechanics is explained by the AdS/CFT correspondence.  

Given this understanding of black holes, it is natural to ask to what extent a similar agreement between gravity and many-body quantum physics is possible in the more realistic case of positive cosmological constant.  We start at an immediate disadvantage however, since as of yet there is no widely accepted holographic description of de Sitter space analogous to AdS/CFT (see \cite{Strominger:2001pn,Maldacena:2002vr,Banks:2006rx,Dong:2010pm,Anninos:2011ui,Coleman:2021nor,Susskind:2022bia,Narovlansky:2023lfz,Tietto:2025oxn} for some attempts).  Nonetheless we can still proceed by doing gravitational calculations in order to gather data that any holographic dual would presumably need to explain.  In this paper that is what we will do.  

\bfig
\includegraphics[height=4cm]{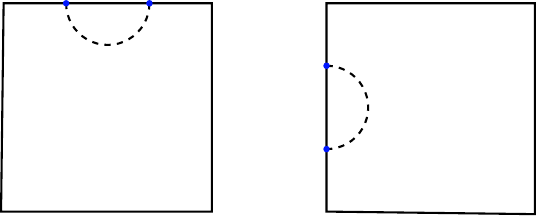}
\caption{Correlation functions in de Sitter space.  On the left are cosmological correlators, as used in computing the fluctuations of the cosmic microwave background.  On the right are observer correlators along the worldline of an observer.}\label{dscorfig}
\efig
Correlation functions in de Sitter space are most often computed at spacelike separation at future infinity.  These are the appropriate correlation functions for explaining the density perturbations that are seen by an observer who lives after a de Sitter-like phase of the universe such as inflation \cite{Mukhanov:1981xt,Hawking:1982cz,Guth:1982ec,Bardeen:1983qw}.  We will call them \textit{cosmological correlators}.  Cosmological correlators are not appropriate however for an observer living in a stable or long-lived de Sitter space, such as the one we seem to currently find ourselves in.   For such an observer it is more appropriate to use correlation functions computed at timelike separation along the worldline of the observer, which we will refer to as \textit{observer correlators}.  The difference between the two kinds of correlators is shown in figure \ref{dscorfig}.  One might hope that observer correlators probe the quantum nature of the de Sitter horizon, giving information about an underlying fundamental description of long-lived or stable de Sitter space.  For example it was recently proposed that there could be a von Neumann algebra of observables acting on some Hilbert space that gives rise to observer correlators \cite{Chandrasekaran:2022cip,Witten:2023xze}; see also \cite{Banks:2013fr} for an earlier version of this idea.  In this paper we study several aspects of the physics of observer correlators.  

The main phenomenon we will analyze in this paper is something we call ``anti-scrambling''.  We will define it more carefully later, but essentially it is a kind of instability of the de Sitter horizon where a perturbation can be amplified by throwing in another perturbation.  This leads to a kind of reverse butterfly effect, where perturbing a state in which a storm will happen next week causes the storm to happen tomorrow instead of preventing it.  Moreover if the time between the two perturbations is large enough, this amplification causes the entire universe to crunch in a cosmological singularity.  Both of these effects arise from the fact that a generic matter perturbation of de Sitter space creates causal contact between opposite sides of the spacetime, which is called the ``tall diagram effect'' \cite{Gao:2000ga,Leblond:2002ns}.

Although the gravitational physics leading to anti-scrambling is quite standard, it poses a serious challenge for any attempt to formulate a fundamental quantum description of the de Sitter horizon.  Ordinary many-body quantum systems exhibit chaos and the butterfly effect, and indeed we will see that anti-scrambling can lead to observer correlators that violate general bounds on quantum expectation values.  What are we to make of this?  We do not have a general answer to this question, but as an empirical observation we will point out that anti-scrambling can be realized in a standard quantum system if we introduce the unusual idea of folded Euclidean time.  Normally Euclidean time can be evolved only forwards and not backwards, due to the operator $e^{\tau H}$ with $\tau>0$ being a bad operator in any quantum system whose energy is unbounded from above.  In a system where the energy is bounded both from above and below however there is no problem with backwards Euclidean evolution, and using this we can evade the bound and produce correlation functions that exhibit anti-scrambling.

The plan of our paper is as follows.  In section \ref{scramblesec} we review black hole scrambling and introduce anti-scrambling in dS using general qualitative arguments.  In section \ref{sec:solution_2D} we construct 2D and 3D de Sitter with matter solutions that quantitatively exhibit anti-scrambling.  In sections \ref{sec:OTOC_1} and \ref{worldlinesec} we compute two-point and four-point observer correlators in 2D and 3D, showing how the out-of-time-order four-point function exhibits antiscrambling.  Finally in section \ref{sec:fold} we explain our proposal to realize anti-scrambling in a standard quantum system by using folded time in Euclidean signature.

Several previous papers have discussed four-point functions on an observer worldline in de Sitter, in particular \cite{Aalsma:2020aib,Kolchmeyer:2024fly,Narovlansky:2025tpb}.  The tall diagram effect was discussed in this context in \cite{Aalsma:2020aib,Narovlansky:2025tpb}.  \cite{Kolchmeyer:2024fly} did not include gravitational effects; as we will discuss in section \ref{sec:solution_2D} the effect they considered is a kinematic recoil effect that has a different origin than the gravitational anti-scrambling we consider here. All these papers found a Lyapunov exponent that is $\frac{4\pi}{\beta}$, while we find an exponent $\frac{2\pi}{\beta}$ saturating the chaos bound.  As this paper was being completed we learned of three other groups working on the same problem, so we coordinated submission.  The other papers are \cite{Chen:2026boh,Cui:2026bcd,Milekhin:2026tbi}.

\section{Overview of scrambling and anti-scrambling}\label{scramblesec}
In this section we review black hole scrambling and contrast it with the behavior of de Sitter horizons, using qualitative arguments based on gravitational focusing.  We will see that de Sitter horizons exhibit an ``anti-scrambling'' effect that is at odds with a straightforward many-body quantum interpretation.  In the following sections we will study this effect more quantitatively in concrete models.  
\subsection{Black hole scrambling}
Black hole horizons are well-known to exhibit scrambling behavior  \cite{Shenker:2013pqa,Roberts:2014ifa,Shenker:2014cwa}.  The basic effect is as follows. We first prepare a state where at time $t_1$ a black hole is going to emit some kind of low-energy particle whose energy is of order the black hole temperature.  We can prepare such a state by acting with a low-energy operator $O_1(t_1)$ on the thermofield double (or equivalently Hartle-Hawking) state $|TFD\ran$.  We then evolve back to some earlier time $t_2<t_1$ and throw in another low-energy particle using an operator $O_2(t_2)$.\footnote{For a big AdS black hole the times $t_1$ and $t_2$ can be thought of as boundary times.  More generally we can define them for example as the proper times when the particles cross a sphere whose radius is twice that of the black hole in the rest frame of the black hole.}  The state of the system after this second particle is
\be
|\psi\ran=O_2(t_2)O_1(t_1)|TFD\ran.
\ee
If $t_1-t_2$ is at most of order the thermal scale $\beta$ then the two particles have no substantial effect on each other, and in particular the first particle is emitted at time $t_1$ as before.  As we continue to increase $t_1-t_2$ however there is a backreaction effect that becomes important: $O(t_2)$ increases the energy of the black hole by $\beta^{-1}$, which causes the horizon to move outwards a little bit.  Indeed by the first law we have
\be
\Delta S=\beta \Delta E \sim 1,
\ee
so from $S=\frac{A}{4G}$ the horizon in Schwarzschild coordinates moves out by
\be
\Delta r\sim \frac{\beta }{S}.
\ee
This may look small, but for $t_1-t_2\gg \beta$ the $O_1$ particle is very close to the horizon when it crosses the $O_2$ particle, so this change can have a big effect on when it is emitted.  Writing the metric as
\be
ds^2=-f(r)dt^2+\frac{dr^2}{f(r)}+r^2d\Omega_{d-2}^2,
\ee
and treating the particles as massless (or just highly boosted), we can find their trajectories by evaluating
\be
(t-t_0)=\int_{r_0}^r\frac{dr'}{f(r')}.
\ee
For the particles in question this integral is dominated by the logarithmic divergence near the horizon, so approximating $f(r)\approx f'(r_h)(r-r_h)$ the two particles cross at a radius
\be
r_{cross}\approx r_h\left(1+e^{-f'(r_h)(t_1-t_2)/2}\right)=r_h\left(1+e^{-\frac{2\pi}{\beta}(t_1-t_2)}\right).
\ee
Here we used the Euclidean regularity requirement
\be
\beta=\frac{4\pi}{f'(r_h)}.
\ee
Thus when 
\be
t_1-t_2\gtrsim \frac{\beta}{2\pi}\log S\equiv t_{scr}
\ee
we have
\be
r_{cross}-r_h\lesssim \Delta r,
\ee
so the outgoing particle is behind the new horizon at the time the backreaction begins and thus is never emitted at all.  See figure \ref{bhshellfig} for an illustration.  This is a gravitational realization of the butterfly effect \cite{Shenker:2013pqa,Roberts:2014ifa,Shenker:2014cwa}: the $O_1$ particle was going to come out later, but then an early perturbation scrambled the system and now it doesn't.

\bfig
\includegraphics[height=4cm]{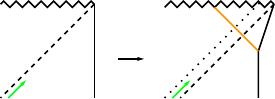}
\caption{Black hole scrambling.  On the left we have a probe particle, shaded green, that will escape from the black hole at some time $t_1$.  If we modify the state by throwing in a massless spherical shell at time $t_2<t_1$ however, then afterwards the asymptotic boundary bends away from the shell (or equivalently the singularity bends down) and the horizon moves outwards (the dotted line indicates where the horizon would have been).  If $t_1-t_2$ is greater than the scrambling time then the green particle is no longer emitted.}\label{bhshellfig}
\efig
The key feature of the black hole horizon that leads to this scrambling effect is that the horizon moves outwards when a particle falls into the black hole.  This is a consequence of the focusing behavior of null geodesics in gravity.  A black hole horizon is defined to be the boundary of the past of future timelike infinity, and the null generators of this boundary are non-expanding going backwards in time by Hawking's area theorem \cite{Hawking:1971tu}. Throwing matter (obeying the null energy condition) through this horizon turns this non-expansion into contraction due to focusing.  This means that prior to the shell crossing the new horizon must lie outside of the old one, since inside the old horizon the transverse area is increasing as we go backwards in time.

\subsection{Anti-scrambling in de Sitter space}
\label{sec:anti_scrambling}
The essential difference between a de Sitter horizon and a black hole horizon is that the transverse area increases as you cross a de Sitter horizon while it decreases as you cross a black hole horizon.  This is because an exterior observer looks at a black hole ``from the outside'', while a de Sitter observer looks at their cosmological horizon ``from the inside''.  In particular this means that the focusing argument of the previous section leads to the opposite conclusion: throwing a particle through a de Sitter horizon causes it to move \textit{away} from the observer rather than towards them.  As a first indication of this phenomenon we can consider the $3+1$ dimensional de Sitter Schwarzschild solution
\be
ds^2=-f(r)dt^2+\frac{dr^2}{f(r)}+r^2d\Omega_{2}^2,
\ee
with
\be
f(r)=1-r^2-\frac{2GM}{r}.
\ee
When $GM\ll 1$, $f$ has two zeros with $r>0$: a black hole horizon at
\be
r\approx 2GM
\ee
and a cosmological horizon at
\be\label{coshor}
r\approx 1-GM.  
\ee
In particular we see that increasing the mass in the static patch decreases the size of the cosmological horizon.  Thus we should expect that if we throw some of this mass through the cosmological horizon, the horizon size should increase and therefore move away.

\bfig
\includegraphics[height=4.5cm]{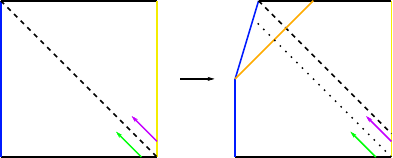}
\caption{Left diagram: de Sitter space with a pair of equal masses at opposite poles of the spatial sphere, shaded blue and yellow.  The dashed line is the horizon for an observer who is part of the left mass.  Right diagram: backreaction on this geometry when the observer radiates a spherical shell (shaded orange) of massless energy.  Their trajectory bends to the right so that their new horizon focuses going backwards in time.   The dotted line shows where their horizon would have been without backreaction.  This backreaction makes the purple particle visible to the observer, and also causes them to see the green particle earlier.}\label{2obsdsfig}
\efig
A simple scenario where we can test this expectation is one where we begin with two objects of equal mass sitting at opposite poles of the spatial de Sitter sphere.\footnote{It is necessary to have both of these masses in order to solve the gravitational constraints, as will become clear when we solve the equations explicitly in the following section.  Outside of the masses the metric is a piece of de Sitter-Schwarzschild, while inside it smoothly caps off at $r=0$.} We will imagine that part of one of these masses is an observer, and that at some time they emit a spherical shell of massless particles.  The Penrose diagram for the backreacted geometry including this shell is shown in figure \ref{2obsdsfig}.  The key point is that, unlike in the black hole case, the trajectory of the observer bends inward after the shell is radiated.  This is because if we evolve the transverse area of their horizon backwards in time, it begins focusing as it crosses the shell.  It thus must reach the mass at the opposite pole (where $r=0$) instead of the past de Sitter infinity (where $r=\infty$).  This means that sending a shell through the horizon allows you to see more of the spacetime, which is the opposite of what happens for black holes.  This is an example of the tall diagram effect, whereby a generic perturbation of de Sitter space that obeys the null energy condition gives any observer causal access to an entire Cauchy slice \cite{Gao:2000ga,Leblond:2002ns} (see also \cite{Batra:2024qju}).

Let's now consider what this effect implies about a possible underlying quantum description of the de Sitter horizon.  Consider a state where in addition to the blue observer and the yellow mass there is also a probe left-moving massless particle that reaches the observer at some proper time $t_1$ (as measured by some clock they are carrying).   Such a particle is shown in green in the left diagram of figure \ref{2obsdsfig}.  The observer will view this particle as having been emitted by their cosmological horizon.  Now suppose that at time $t_2<t_1$ the observer generates a massless outgoing shell, as in the right diagram of figure \ref{2obsdsfig}.  Since this causes their trajectory to bend to the right, they now encounter the green particle at a new proper time $t_1'<t_1$.  This is what we mean by anti-scrambling: rather than delaying or preventing the appearance of the green particle as in the black hole case, in de Sitter space the shell causes the horizon to emit the green particle earlier.\footnote{The ``anti-'' here does not mean that the system does not scramble in the sense of being free or integrable.  We think of anti-scrambling as being a different kind of scrambling, analogous to ``anti-matter'' or ``anti-de Sitter space''.}  In other words, in de Sitter space when a butterfly flaps its wings it makes the storm come sooner.  In fact it does more: it can expose the observer to ``storms'' such as the purple particle which otherwise would never have arrived at all!

\section{Low-dimensional gravitational solutions}
\label{sec:solution_2D}
We now give a more quantitative discussion of anti-scrambling by presenting solutions of 2d and 3d gravity that exhibit it.  

\subsection{Reduction from 3d to 2d}
We begin with three-dimensional Einstein gravity with positive cosmological constant, in units where the vacuum energy $\rho_0$ obeys
\be
8\pi G\rho_0=1
\ee
so that the de Sitter radius is one.  A general spherically-symmetric metric also invariant under the reflection $\phi'=2\pi-\phi$ is
\be
ds^2=\wt{g}_{\mu\nu}(x)dx^\mu dx^\nu+r(x)^2d\phi^2,
\ee
where $\phi$ has periodicity $2\pi$ and $\wt{g}_{\mu\nu}$ is an arbitrary two-dimensional Lorentzian metric. Spherical and reflection symmetry also imply that the energy-momentum tensor has the form
\be
T=\wt{T}_{\mu\nu}dx^\mu dx^\nu+T_{\phi\phi}d\phi^2.
\ee
The gravitational equations of motion for this system are
\begin{align}\nonumber
-\wt{\nabla}_\mu \wt{\nabla}_\nu r+\wt{g}_{\mu\nu}\left(\wt{\nabla}^2r+r\right)&=8\pi G r \wt{T}_{\mu\nu}\\
\wt{R}-2&=-16\pi G r^{-2}T_{\phi\phi}.
\end{align}
On any timelike locus in the two-dimensional geometry where $r(x)=0$, regularity of the three-dimensional geometry requires that
\be
n^\mu \partial_\mu r=1
\ee
where $n^{\mu}$ is the inward-pointing normal vector to the locus in the two-dimensional Lorentzian geometry.  If there is a matter particle of mass $M$ on this locus, then this boundary condition is instead 
\be
n^\mu \partial_\mu r=1-4GM
\ee
due to a conical deficit $\delta=8\pi G M$.  In particular we require that $4GM<1$ so that the spacetime is not completely closed off.  We will consider spacetimes where the three-dimensional topology is $\mathbb{S}^2\times \mathbb{R}$, so the two-dimensional geometry is a spatial interval between two $r=0$ boundaries as in figure \ref{2obsdsfig}.  We will focus on ``dust-like'' matter where the transverse pressure $T_{\phi\phi}$ vanishes, in which case after the redefinitions
\begin{align}\nonumber
T_{\mu\nu}&\equiv 2\pi r \wt{T}_{\mu\nu}\\
\Phi&\equiv \frac{r}{4G}\label{2d3d}
\end{align}
we arrive at the equations of motion for Jackiw-Teitelboim gravity with positive cosmological constant coupled to matter:
\begin{align}\nonumber
\nabla_\mu\nabla_\nu\Phi+g_{\mu\nu}\left(\nabla^2\Phi+\Phi\right)&=T_{\mu\nu}\\
R&=2.
\end{align}
Here we have dropped the tildes on the two-dimensional metric and covariant derivatives.  The boundary condition at $r=0$ in JT language becomes\footnote{This boundary condition was previously considered in \cite{Goel:2020yxl}, with a rather different interpretation.} 
\begin{align}\nonumber
\Phi&=0\\
n^\mu \nabla_\mu\Phi&=\frac{1}{4G}\left(1-4 GM\right).\label{JTBC}
\end{align}
A two-dimensional action that leads to these equations of motion and boundary conditions is
\be
S=\frac{1}{2}\int d^2 x \sqrt{-g}\,\Phi(R-2)+\frac{1}{4 G}\int_{\Gamma} dx \sqrt{-h}(1-4GM)+S_{matter},
\label{action1}
\ee
where $\Gamma$ is the spatial boundary and $h$ is the induced metric on $\Gamma$.  This action is not quite the one we will use however; we need to promote the boundary mass parameter $M$ to a field $\M(x)$ to allow it to be different on different boundaries and also change if particles are emitted or absorbed from the boundary.  We postpone the full action including dynamical $\M$ to section \ref{worldlinesec} below; for now we will just allow $M$ to change when needed to allow particle emission/absorption.  Using \eqref{2d3d} we have a precise equivalence between solutions of $2+1$ dimensional Einstein gravity with $\Lambda>0$ and spherical dust and solutions of JT gravity with $\Lambda>0$ and arbitrary matter.\footnote{A similar map was constructed between spherical dust perturbations of the BTZ black hole and AdS JT gravity in \cite{Harlow:2021dfp}, and the solution method we use here is adapted from there.}  We therefore can use either language in discussing the solutions of these equations; for the most part we will use the two-dimensional language since it is a bit simpler.  

\subsection{Solution method}
By now there are a number of standard methods for solving the JT equations of motion in the presence of matter, so we will leave some details to appendix \ref{JTapp}.  We work in Kruskal coordinates where the metric is given by
\be\label{kruskal}
ds^2=-\frac{4\,dX^+dX^-}{(1-X^+X^-)^2},
\ee
where as usual we have
\be
dX^+dX^-\equiv \frac{1}{2}(dX^+\otimes dX^-+dX^-\otimes dX^+).
\ee
These coordinates are related to the embedding coordinates where $dS_2$ is realized as a surface 
\be
-T^2+Y^2+Z^2=1
\ee
in three dimensional Minkowski space by
\begin{align}\nonumber
T\pm Z&=\frac{2X^\pm}{1-X^+X^-}\\
Y&=\frac{1+X^+X^-}{1-X^+X^-}.
\end{align}
The general dilaton solution away from any matter is (see for example \cite{Alonso-Monsalve:2024oii}) 
\be\label{dilsol}
\Phi=\frac{a_+X^++a_-X^-+b(1+X^+X^-)}{1-X^+X^-}.
\ee
These solutions are characterized by the constant quantity
\be
\kappa\equiv \Phi^2+g^{\mu\nu}\nabla_\mu\Phi\nabla_\nu\Phi=b^2-a_+a_-.
\ee
Near a $\Phi=0$  boundary this quantity is determined by the boundary conditions \eqref{JTBC} to be
\be\label{kappabc}
\kappa=\frac{(1-4GM)^2}{16G^2}.
\ee
In particular the $dS_3$ solution with equal masses at the spatial poles shown on the left side of figure \ref{2obsdsfig} corresponds to a dilaton solution with 
\begin{align}\nonumber
a_\pm&=0\\
b&=\frac{1-4GM}{4G}.\label{0sol}
\end{align}
For this solution the boundary trajectories as a function of proper time $t$ are given by
\be\label{L0}
X^\pm=\mp e^{\mp t}
\ee
for the left boundary and
\be
X^\pm=\pm e^{\pm t}
\ee
for the right boundary.
Other solutions with the same value of $\kappa$ are related to this one by acting with $dS_2$ isometries.  If the masses are not equal then there is no solution since we cannot pick a consistent value of $\kappa$, which is an illustration of the power of the gravitational constraints in a closed universe.  Solutions with $\kappa<0$ are also interesting, up to a $dS_2$ isometry they can be taken to have $a_+=a_-=a$ and $b=0$.  For this solution the dilaton vanishes on the time slice $X^++X^-=0$, which we can think of as a big bang or big crunch depending on whether we keep the part of the geometry to the future or past of this time slice.  In \cite{Alonso-Monsalve:2024oii} solutions with $\kappa>0$ were called spacelike and solutions with $\kappa<0$ were called timelike; there are also null solutions with $\kappa=0$ that will not play a role in this paper.\footnote{In \cite{Alonso-Monsalve:2024oii} JT gravity with positive cosmological constant was studied using periodic spatial boundary conditions.  These give a model of the near-extremal higher dimensional de Sitter black hole, while the $\Phi=0$ boundaries we use here give a model of near-empty higher dimensional de Sitter.}

We will take the matter in our solutions (except for the mass at the boundaries, which is incorporated by the boundary conditions) to consist of null shells of the form
\be\label{Tnull}
T_{\mp\mp}(X)=k\delta(X^\mp-X^\mp_0).
\ee
Here the upper/lower sign describes a right/left moving shell.  The derivative of the dilaton has a jump across such shells, which we show in appendix \ref{JTapp} leads to the discontinuities
\begin{align}\nonumber
\Delta a_\pm&=-k(X_0^\mp)^2\\\nonumber
\Delta a_\mp&=-k\\
\Delta b&=kX_0^\mp\label{Dabc}
\end{align}
in the dilaton solution.  Here $\Delta$ means the jump from past to future across the shell.  In three dimensional language these are the Barrabes-Israel jump conditions describing the jump of the transverse extrinsic curvature across the shell \cite{Barrabes:1991ng}.  If one of these shells intersects either boundary then its energy in the rest frame of the boundary is given by
\be\label{shellE}
\omega=k\dot{X}^{\mp},
\ee
where $X^\pm(t)$ is the boundary trajectory as a function of proper time and again the upper/lower sign is for a right/left moving shell.  Using \eqref{kappabc}, \eqref{Dabc}, and \eqref{shellE} we can construct the dilaton solution for an arbitrary number of shells emitted/absorbed/reflected from the boundaries or propagating to/from the de Sitter future/past infinities.   

\subsection{One shell}
The first nontrivial solution we will consider describes a situation where an observer at the left boundary emits a spherical shell of energy $\omega$ in their rest frame at $t=0$.  We can think of this as a quantum state
\be
O(0)|M\ran,
\ee
where $|M\ran$ is the Hartle-Hawking state conditioned to have a pair of masses each of mass $M$ on opposite sides of de Sitter space and $O(0)$ creates a shell of energy $\omega$ emitted from the left mass at proper time $t=0$.  The Penrose diagram of this solution is shown in the right diagram of figure \ref{2obsdsfig}.  The solution to the past of the shell is  just \eqref{0sol}.  Defining
\be\label{b0def}
b_0\equiv \frac{1-4GM}{4G},
\ee
the solution to the future of the shell is
\begin{align}\nonumber
a_\pm&=-\omega\\
b&=b_0+\omega
\end{align}
and the trajectory of the left boundary to the future of the shell is
\begin{align}\nonumber
X^+(t)&=\frac{-\cosh \frac{t}{2}+\sqrt{1+2\omega/b_0}\sinh \frac{t}{2}}{\cosh \frac{t}{2}+\sqrt{1+2\omega/b_0}\sinh \frac{t}{2}}\\
X^-(t)&=\frac{\sinh\frac{t}{2}+\sqrt{1+2\omega/b_0}\cosh \frac{t}{2}}{-\sinh\frac{t}{2}+\sqrt{1+2\omega/b_0}\cosh \frac{t}{2}}.
\label{eq:JT_trajectory}
\end{align}
Assuming that the initial boundary mass is not close to $\frac{1}{4G}$, $b_0$ is of order the de Sitter entropy
\be
S_{dS}=\frac{\pi}{2G}
\ee
so if $\omega$ is of order the de Sitter temperature then we can think of $\frac{\omega}{b_0}$ as a small quantity.  In fact this is as good a place as any to mention that the Euclidean saddle point for the partition function of this theory is a half-sphere with action
\be
S_E=-2\pi b_0,
\ee
so the thermodynamic entropy $-S_E$ at fixed $M$ is just given by $2\pi b_0$.

Now consider a left-moving probe particle which, in the absence of the shell, would arrive at the left boundary at proper time $t_1$.  Such a particle is shown in green in figure \ref{2obsdsfig}.  From \eqref{L0} it lies at $X^+=-e^{-t_1}$.  We can therefore determine the proper time $t_1'$ at which it arrives at the backreacted left boundary by solving
\be\label{tadv}
-e^{-t_1}=\frac{-\cosh \frac{t_1'}{2}+\sqrt{1+2\omega/b_0}\sinh \frac{t_1'}{2}}{\cosh \frac{t_1'}{2}+\sqrt{1+2\omega/b_0}\sinh \frac{t_1'}{2}}.
\ee
In particular if $t_1=\infty$ then we have
\be
t_1'=\log \left(\frac{\sqrt{1+\frac{2\omega}{b_0}}+1}{\sqrt{1+\frac{2\omega}{b_0}}-1}\right)\equiv t_\infty,
\ee
which we can approximate at small $\omega/b_0$ as
\be
t_\infty = \log \left(\frac{2b_0}{\omega}\right)+O\left(\frac{\omega}{b_0}\right)=t_{scr}+O(1).
\ee
Here
\be
t_{scr}\equiv \log S_{dS}
\ee
is the de Sitter scrambling time.  Thus if the probe particle would have arrived at all without the matter shell, with the shell it arrives within a scrambling time!  How much earlier depends on the scale of $t_1$: if $t_1\ll t_{scr}$ we have\footnote{This formula is actually true to all orders in $\frac{\omega}{b_0} e^{t_1}$, it is valid until $\frac{\omega^2}{b_0^2} e^{t_1}\sim 1$ which happens at twice the scrambling time.}
\be\label{tadv1}
t_1'\approx t_1-\frac{\omega}{b_0}\sinh t_1,
\ee
while if $t_1\gg t_{scr}$ we have
\be
t_1'\approx t_\infty-\frac{2b_0}{\omega}e^{-t_1}.
\ee
In both cases we have suppressed terms that are higher order in $\omega/b_0$, and in both cases there is an anti-scrambling time advance instead of the time delay familiar from black holes.

\subsection{Two shells}\label{2shellsec}
\bfig
\includegraphics[height=4cm]{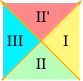}
\caption{Spacetime regions for two colliding shells.  According to the blue observer, region I is to the exterior of both shells, region III is inside both shells, and region II' is the post-collision region.}\label{collisionfig}
\efig
We now consider a solution with two shells.  The idea is to describe the geometry of a state
\be
|\psi\ran=O_2(t_2)O_1(0)|M\ran,
\ee
where $|M\ran$ is the Hartle-Hawking state conditioned to have a pair of masses each of mass $M$ on opposite sides of de Sitter space, $O_1(0)$ is an operator that creates a shell of energy $\omega_1$ that is absorbed by the left mass at proper time $t=0$, and $O_2(t_2)$ is an operator that emits a shell of energy $\omega_1$ at time $t_2<0$ from the left mass.  These two shells collide somewhere in the spacetime, and the four spacetime regions relative to this collision are shown in figure \ref{collisionfig}.  Spacelike to $O_2(t_2)$ the geometry is the same as in the state $O_1(0)|M\ran$, which is the time-reverse of the state we considered in the previous section.  Thus the solution in region I is \eqref{0sol} and the solution in region II is 
\begin{align}\nonumber
a_{\I\I,\pm}&=\omega_1\\
b_{\I\I}&=b_0+\omega_1,
\end{align}
with boundary trajectory
\begin{align}\nonumber
X^-_{\I\I}(t)&=\frac{\cosh\frac{t}{2}+\sqrt{1+\frac{2\omega_1}{b_0}}\sinh \frac{t}{2}}{\cosh\frac{t}{2}-\sqrt{1+\frac{2\omega_1}{b_0}}\sinh\frac{t}{2}}\\
X^+_{\I\I}(t)&=\frac{\sinh\frac{t}{2}-\sqrt{1+\frac{2\omega_1}{b_0}}\cosh\frac{t}{2}}{\sinh\frac{t}{2}+\sqrt{1+\frac{2\omega_1}{b_0}}\cosh \frac{t}{2}}.
\end{align}
The location of the right-moving shell created by $O_2(t_2)$ is
\be
X_2^-=X_{\I\I}^-(t_2),
\ee
and its momentum $k_2$ is given by
\be
k_2=\frac{\omega_2}{\dot{X}_{\I\I}(t_2)}.
\ee

\bfig
\includegraphics[height=5cm]{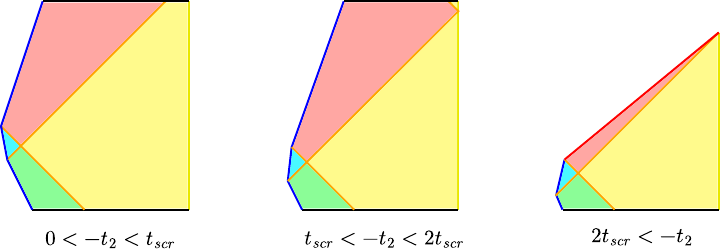}
\caption{Backreacted geometry with two shells, as a function of the time separation $-t_2$.  The crunch singularity in region II' for sufficiently early $O_2$ is indicated by a red line.}\label{2shellfig}
\efig
What happens after the collision depends on the magnitude of $t_2$, with the possible options shown in figure \ref{2shellfig}.  As we increase $-t_2$ the first qualitative change of the solution happens when $X_2^-$ becomes negative, so that the shell created by $O_2(t_2)$ reaches the right boundary.  This happens when
\be
\tanh \frac{t_2}{2}=-\frac{1}{\sqrt{1+\frac{2\omega_1}{b}}},
\ee
or in other words when
\be
-t_2=t_{scr}+O(1).
\ee

To understand the next qualitative change we need to compute the invariant $\kappa$ in regions II' and III.  
In region II' we have
\begin{align}\nonumber
a_{\I\I',+}&=-k_2 (X_2^-)^2\\\nonumber
a_{\I\I',-}&=-k_2\\
b_{\I\I'}&=b_0+k_2X_2^-,
\end{align}
while in region III we have
\begin{align}\nonumber
a_{\I\I\I,+}&=\omega_1-k_2 (X_2^-)^2\\\nonumber
a_{\I\I\I,-}&=\omega_1-k_2\\
b_{\I\I\I}&=b_0+\omega_1+k_2X_2^-.
\end{align}
After some simplification we then have
\begin{align}\nonumber
\kappa_{\I}&=b_0^2\\\nonumber
\kappa_{\I\I}&=b_0^2+2b_0\omega_1\\\nonumber
\kappa_{\I\I'}&=b_0^2+\frac{2b_0\omega_2}{\sqrt{1+\frac{2\omega_1}{b_0}}}\left(1-\frac{\omega_1}{b_0}\big(\cosh t_2-1\big)\right)\\
\kappa_{\I\I\I}&=b_0^2+2b_0\left(\omega_1+\omega_2\sqrt{1+\frac{2\omega_1}{b_0}}\right).
\end{align}
For small $\frac{\omega_1}{b_0}$ and $\frac{\omega_2}{b_0}$ we have $\kappa_{\I\I}\approx \kappa_{\I\I\I}\approx \kappa_{\I}$ for all $t_2$, so the backreaction effects are small.  $\kappa_{II'}$ however has a term that grows exponentially with time, and when
\be
-t_2=2t_{scr}+O(1)
\ee
it changes sign so the dilaton solution becomes timelike.  This means that the surface $\Phi=0$ becomes spacelike, leading to a big crunch singularity in region II$^\prime$.  The seemingly innocent operation of dropping a thermal-scale particle through a de Sitter horizon that will emit a thermal-scale particle two scrambling times later has destroyed the universe!  This is an extreme form of anti-scrambling, and as far as we know there is no analogous phenomenon for an exterior observer of a black hole.

\subsection{Localized perturbation: recoil effect}\label{sec:recoil}
So far we only discussed spherically-symmetric matter in $dS_3$ with vanishing transverse pressure, which we showed is equivalent to our 2D JT model coupled to matter. In $dS_3$ however our observer could also send their outgoing perturbation in a particular direction, breaking spherical symmetry. Aside from the anti-scrambling effect we just studied, this also causes the observer to recoil in the direction opposite to the perturbation. In this subsection we describe this effect quantitatively. 

We will model an observer with mass $M$ as a conical defect in $dS_3$, with metric given by
\begin{equation}
\begin{aligned}
    ds^2 =\,& -(1-r^2)dt^2+\frac{dr^2}{1-r^2}+r_0^2r^2d\phi^2\\
    =\,&-\frac{4dx^+dx^-}{(1-x^+x^-)^2}+r_0^2\qty(\frac{1+x^+x^-}{1-x^+x^-})^2d\phi^2
\end{aligned}
\label{eq:metric_background}
\end{equation}
with $r_0 = 1-4GM$, $\phi\in[-\pi, \pi]$.  We are using lower-case $x^\pm$ here because, once the localized perturbation is included, these coordinates are related to the ones in the previous subsections by a discontinuous diffeomorphism.  At time $t_w$, the observer throws a localized perturbation with energy $\omega$ toward the horizon along the $\phi=0$ direction, with stress-energy tensor given by
\begin{align}
	T_{--} = \frac{\omega}{r_0} e^{-t_w}\delta(x^-)\delta(\phi).
\end{align}
We use a shockwave ansatz for the perturbed metric:
\begin{align}
	ds^2 =\,& -\frac{4 dx^+dx^-}{(1-x^+x^-)^2}dx^+ dx^-+\frac{4 \delta(x^-)h(\phi)(dx^-)^2}{(1-x^+x^-)^2}+r_0^2\qty(\frac{1+x^+x^-}{1-x^+x^-})^2d\phi^2
\end{align}
Solving Einstein's equation:
\begin{align}
	8\pi G T_{--} = -2\delta(x^-)\qty(\frac{h''(\phi)}{r_0^2}+h(\phi)) = 8\pi G \frac{\omega}{r_0} e^{- t_w}\delta(x^-)\delta(\phi),
\end{align}
we obtain
\begin{align}\label{eq:shift_local}
	h(\phi) =\ & -\frac{2\pi  G_N \omega}{\sin(r_0\pi)} e^{- t_w}\cos[r_0(\pi-|\phi|)].
\end{align}
We emphasize that the existence of this solution requires $M>0$: this is because otherwise the conservation of energy does not allow the emission of a shell, and the conservation of the energy-momentum tensor is a consequence of Einstein's equation.  

Here is one way to understand the physical meaning of $h(\phi)$.  Imagine there is an incoming probe particle from angular direction $\phi$ that, in the absence of the emitted perturbation, would reach the observer at $t=0$.  Now the observer emits their localized perturbation towards $\phi=0$ at time $t_w<0$, causing them to recoil in the $\phi=\pi$ direction. Depending on $\phi$, the incoming probe particle can reach the observer sooner or later. Letting $t'(\phi)$ be the time the probe particle actually arrives, we have
\begin{align}
    t'(\phi) = -\log(1-h(\phi)).\label{eq:time_localized}
\end{align}
Equation \ref{eq:time_localized} is the localized analog of \eqref{tadv}. When $h(\phi)<0$, the observer will receive the probe signal earlier, while when $h(\phi)>0$, the observer will receive the probe signal later. When $h(\phi)>1$, the observer will miss the signal. \\

\begin{figure}
    \centering
    \includegraphics[height=6cm]{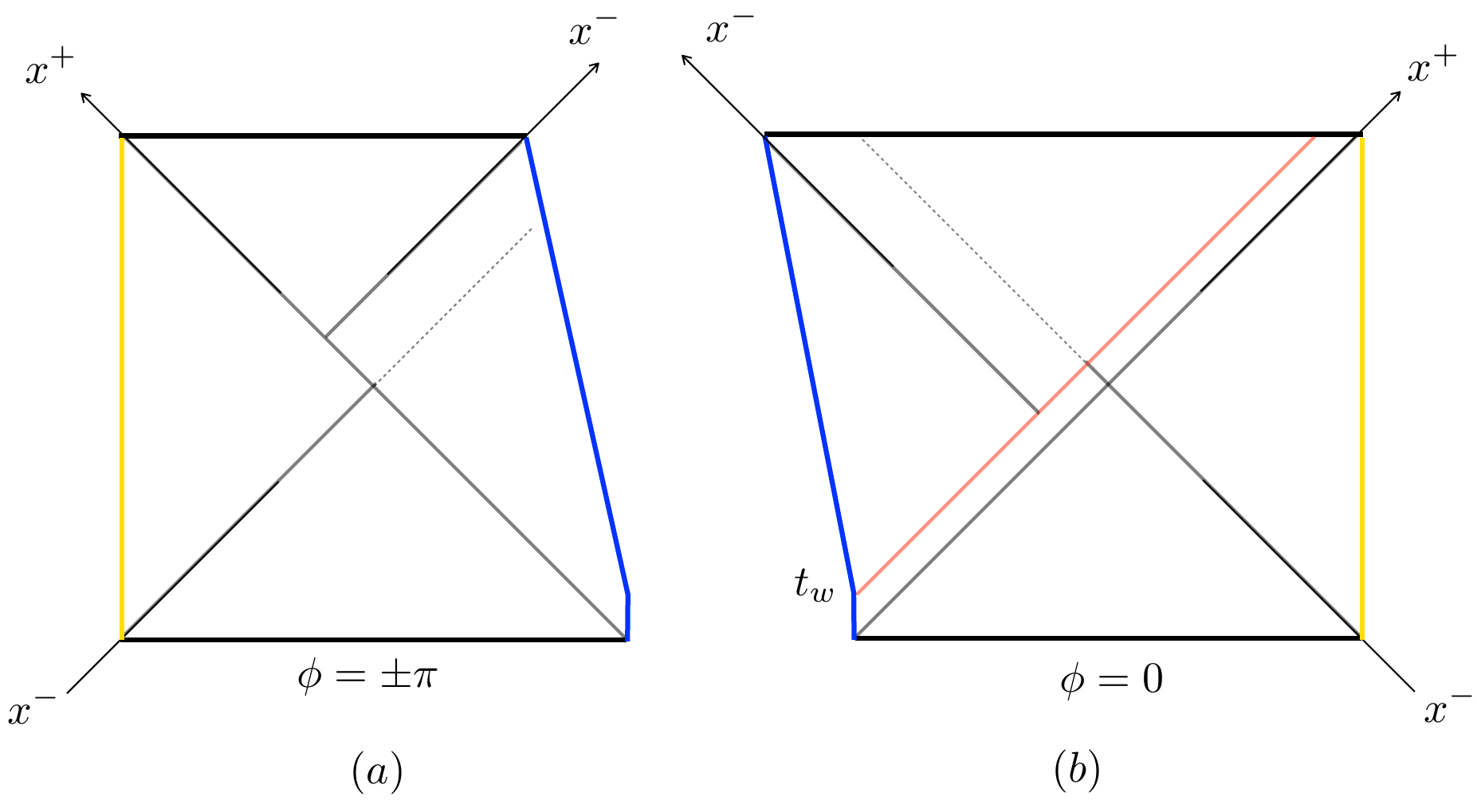}
    \caption{(b): $\phi=0$ section of the Penrose diagram; (a): $\phi = \pm\pi$ section of the Penrose diagram with the left and right sides reflected.}
    \label{fig:recoil}
\end{figure}
We can study three different regimes:
\begin{itemize}
    \item Light observer: $G M\ll 1$.
    In this regime
    \begin{align}
        h(\phi)\approx \frac{\omega}{2M}e^{-t_w}\cos(r_0\phi),\label{eq:shift_recoil}
    \end{align}
where we used $1-r_0 = 4GM$.
In \eqref{eq:shift_recoil}, the prefactor in front of the exponential is given by $\frac{\omega}{M}$ and $G$ disappears. The shift becomes of order one when $-t_w = t_M^* = \log(\frac{2M}{\omega})$. This is a purely kinematic effect due to the recoil just mentioned; it has nothing to do with gravity.  The sign of the shift depends on $\phi$, since as just explained the observer can recoil either towards or away from the incoming probe particle. Indeed in the direction of the perturbation $h(\phi=0)>0$ (figure \ref{fig:recoil}(b)), while along the opposite direction $h(\phi = \pm\pi)<0$ (figure \ref{fig:recoil}(a)). In figure \ref{fig:recoil}(a) we draw the cross-section of $\phi = \pi$ with left-right side of the Penrose diagram reflected. If we paste (a) and (b) together along the observer's worldlines, we can consider it as a cross-section of $dS_2$ geometry and the recoil effect here is what was called scrambling in \cite{Kolchmeyer:2024fly}.

\item Heavy observer: $M$ is of order the Planck mass, $r_0\ll 1$ but $\frac{r_0}{G }\gg 1$. 
\begin{align}
    h(\phi)\approx - \frac{2 G \omega}{r_0} e^{-t_w}
\end{align}
In this regime, the angular dependence is weak and $h(\phi)$ is negative. We recover the anti-scrambling effect discussed in earlier sections. The scrambling time is given by $t_{scr} = \log(S)$ where $S = \log(\frac{\pi r_0}{2G})$ is the entropy of the cosmological horizon. 

\item Intermediate regime

In the intermediate regime, $M\sim\frac{1}{G_N}$. $\sin(r_0\pi)$ is positive and of order $1$. Recoil and anti-scrambling effects are both present in equation \eqref{eq:shift_local} and cannot be clearly separated. 
\end{itemize}
For the remainder of this paper we will stick to spherically-symmetric perturbations, since as far as we can tell the recoil effect is non-gravitational in nature and tells us about how the observer sees the horizon rather than what the horizon is doing.

\section{Observer correlators I: classical gravity}
\label{sec:OTOC_1}
We now turn to studying the impact of anti-scrambling on observer correlators.  As a concrete model we will consider a pair of massless scalar fields $\chi_1$ and $\chi_2$ coupled to JT gravity via the action \eqref{action1} with 
\be
S_{matter}=-\frac{1}{2}\int d^2 x \sqrt{-g}\partial_\mu \chi_i \partial^\mu \chi_i.
\label{eq:action_2D_matter}
\ee
We are interested in correlation functions of the form
\be
\lan M|O_n[\chi]\ldots O_1[\chi]|M\ran,
\ee
where $|M\ran$ is again the Hartle-Hawking state conditioned on the presence of the boundary masses $M$ and $O_m[\chi]$ are local functionals of $\chi_i$ evaluated on the left boundary as a function of proper time.  In this section we will essentially treat gravitational interactions classically, postponing quantum gravity effects to the next section. In order for the theory to be well-defined we need boundary conditions for the scalar fields, so we will adopt Dirichlet boundary conditions at both boundaries. 

At zeroth order in $G$ we are studying the correlation functions of a free scalar field in a pure de Sitter background, which we will again study using Kruskal coordinates
\be
ds^2=-\frac{4dX^+dX^-}{(1-X^+X^-)^2}.
\ee
For now we will put all operators on the left boundary.  The massless wave equation in this geometry is
\be
\partial_+\partial_-\chi=0,
\ee
and a convenient pair of complete sets of modes is
\be
f_{\pm,p_{\mp}}(X^+,X^-)=\frac{1}{\sqrt{-2p_{\mp}}}\left(e^{ip_\mp X^\mp}-e^{-i\frac{p_\mp}{ X^\pm}}\right)
\ee
with $p_{\mp}<0$.  The second term is necessary to obey the Dirichlet boundary condition
\be
f_{\pm,p_{\mp}}(X^+,-1/X^+)=0.
\ee
We emphasize that $f_{+,p_-}$ and $f_{-,p_{+}}$ are each separately a complete set of modes.  The prefactor is chosen so that these modes are normalized in the Klein-Gordon inner product; this is most easily checked by integrating on the horizon,\footnote{Because of the singularity in the second term, it is safest to evaluate this integral as the limit of an integral at $X^\pm=\mp \epsilon$ as $\epsilon\to 0^+$.  Strictly speaking this is not a Cauchy slice, but we can make it one by taking the union with the surface $X^{\mp}=\mp 1/\epsilon$.  One can check that the integral over the latter surface vanishes as $\epsilon \to 0^+$.}
\be
\lan f_{\pm,p'_{\mp}},f_{\pm,p_{\mp}}\ran_{KG}\equiv i\int_{X^\pm=0} dX^{\mp}\left(f_{\pm,p'_{\mp}}^*\partial_\mp f_{\pm,p_{\mp}}-\partial_\mp f_{\pm,p'_{\mp}}^* f_{\pm,p_{\mp}}\right)=2\pi \delta(p_\mp-p_{\mp}').
\ee
The $f_+$ modes are natural for describing right-moving particles near the left boundary, while the $f_-$ modes are most natural for describing left-moving particles.  

Since we are using Dirichlet boundary conditions we cannot directly compute correlation functions of $\chi$ on the observer worldline, so we instead compute correlation functions of its normal derivative.  Using the normal vector
\be
n=-X^+\partial_+-X^-\partial_-,
\ee
we have
\be
\partial_n\chi_i=\int_{-\infty}^0 \frac{dp_{\mp}}{2\pi}\left(g_{\pm,p_\mp}a_{\pm,p_\mp}+g_{\pm,p_\mp}^*a_{\pm,p_\mp}^\dagger\right)
\ee
with
\begin{align}\nonumber
g_{\pm,p_\mp}&=\frac{i}{\sqrt{2}}\sqrt{-p_{\mp}}\left(X^\mp e^{ip_\mp X^{\mp}}-\frac{1}{X^{\pm}}e^{-i\frac{p_{\mp}}{X^\pm}}\right)\\
&=\pm 2i\sqrt{p^\pm}e^{\pm t}e^{\mp 2ip^{\pm}e^{\pm t}}.\label{gpm}
\end{align}
In the second line we evaluated the mode at the left boundary
\be\label{bound0}
X_\pm(t)=\mp e^{\mp t},
\ee
and also introduced a raised momentum
\be
p^\pm\equiv-\frac{p_\mp}{2}
\ee
that is a bit more intuitive since it is positive and has a sign that aligns with the direction the mode is moving.  $a_{\pm,p_{\mp}}$ is the annihilation operator for the mode $f_{\pm,p_{\mp}}$, normalized so that
\be
[a_{\pm,p_{\mp}},a_{\pm,p_{\mp}'}^\dagger]=2\pi \delta(p_{\mp}-p_{\mp}').
\ee

As a warmup we can compute the Wightman two-point function of $\partial_n \chi$:
\be
G_2(t_1,t_2)\equiv \lan M|\partial_n\chi_i(t_2)\partial_n\chi_i(t_1)|M\ran= \int_{-\infty}^0\frac{dp_\mp}{2\pi}g_{\pm,p_\mp}(t_2)g_{\pm,p_{\mp}}(t_1)^*.
\ee
We do the computation with both sets of modes as a cross-check. Using \eqref{gpm} we have
\begin{align}\nonumber
G_2 &=\frac{4}{\pi}e^{\pm (t_2+t_1)}\int_0^\infty dp^\pm p^\pm e^{\mp 2ip^\pm\left(e^{\pm t_2}-e^{\pm t_1}\right)}\\
&=-\frac{1}{4\pi} \frac{1}{\sinh^2\left(\frac{t_2-t_1}{2}\right)}.
\end{align}
The convergence of the integral is distributional for real $t_1$, $t_2$, we can make it genuinely convergent by introducing the Wightman prescription
\be
t_k=\wt{t}_k-i\epsilon k\label{ieps}
\ee
and slightly rotating the contour.  The main point to notice is that the analytic continuation of this correlation function is periodic in imaginary time, with periodicity $2\pi$.  This is the usual statement that field theory correlators in de Sitter space behave as if the fields are in equilibrium with a horizon of temperature $\frac{1}{2\pi}$.  

Turning now to the four-point function, following \cite{Shenker:2014cwa} to diagnose anti-scrambling we should compute\footnote{We have used two species of field to avoid distracting extra two-point contributions.}
\be\label{G4def}
G_4(t_1,t_2,t_3,t_4)\equiv \lan M|\partial_n \chi_2(t_4)\partial_n\chi_1(t_3)\partial_n\chi_2(t_2)\partial_n\chi_1(t_1)|M\ran
\ee
in the regime where $t_1,t_3$ are similar, $t_2,t_4$ are similar, and
\be\label{larget}
t_1-t_2\gg 1.
\ee
  At zeroth order in $G$ this correlator is just a product of the two-point functions we just computed, so we are interested in the first gravitational correction.  There are two possible kinds of gravitational corrections: corrections to two-point functions, which do not grow with $t_1-t_2$, and corrections arising from gravitational scattering, which can grow.  When \eqref{larget} holds the dominant contribution is from $2-2$ gravitational scattering in the eikonal regime \cite{Shenker:2014cwa}, so we can write the four-point function as
\be
G_4\approx\int_{-\infty}^0\frac{dp_-}{2\pi}\int_{-\infty}^0\frac{dp_+}{2\pi}g_{+,p_-}(t_4)g_{+,p_-}(t_2)^*g_{-,p_+}(t_3)g_{-,p_+}(t_1)^* e^{i\delta(p_+p_-)},
\ee
where $e^{i\delta}$ is the $2-2$ gravitational scattering phase.   By Lorentz invariance it depends only on the combination $p_+p_-$.  The bulk geometry describing the scattering is the two-shell geometry shown in figure \ref{2shellfig}.  The derivative of the scattering phase is related to the time advance of the outgoing particles in the usual way, in particular for particle one we have\footnote{The logic of this formula is to form a narrow wave packet and then use the stationary-phase approximation to see how the phase shift causes the peak of the wave packet to move.}
\be\label{advdelt}
X^+_1-(X_1^+)'=\frac{\partial \delta}{\partial p_+},
\ee
so using \eqref{tadv1} and \eqref{bound0}, and also 
\be
\omega_2=-p_-e^{t_2},
\ee
we have
\be
\delta'\approx -\frac{1}{b_0}e^{t_2-t_1}\sinh \left(t_1-t_2\right)\approx -\frac{1}{2b_0}
\ee
and thus
\be
\delta \approx -\frac{p_+p_-}{2b_0}=-\frac{2p^+p^-}{b_0}.
\label{eq:phase}
\ee
Here we used that there is no gravitational scattering at zero energy, and in getting rid of the time-dependence we also used \eqref{larget}.  Using \eqref{gpm} we therefore have
\begin{align}\nonumber
G_4&=\frac{16}{\pi^2}e^{-t_1-t_3+t_4+t_2}\int_0^\infty dp^+\int_0^\infty dp^-p^+p^-e^{-2ip^+\left(e^{t_4}-e^{t_2}\right)-2ip^-\left(e^{-t_1}-e^{-t_3}\right)-2i\frac{p^+p^-}{b_0}}\\
&=\frac{16b_0^2}{\pi^2}e^{-t_1-t_3+t_4+t_2}I(\beta,\gamma)\label{G41}
\end{align}
where
\be
I(\beta,\gamma)\equiv\int_0^\infty dx \int_0^\infty dy \,xy\,e^{-i\beta x-i\gamma y-2ix y}
\label{eq:I}
\ee
with
\begin{align}\nonumber
\beta&\equiv 2\sqrt{b_0}\left(e^{t_4}-e^{t_2}\right)=4\sqrt{b_0}e^{\frac{t_4+t_2}{2}}\sinh\left(\frac{t_4-t_2}{2}\right)\\
\gamma&\equiv 2\sqrt{b_0}\left(e^{-t_1}-e^{-t_3}\right)=4\sqrt{b_0}e^{-\frac{t_1+t_3}{2}}\sinh\left(\frac{t_3-t_1}{2}\right).\label{bgdef}
\end{align}
This integral is again only distributionally convergent, we can render it actually convergent by using the Wightman prescription \eqref{ieps} and slightly rotating the contours.  We can simplify it by using differentiation under the integral sign to remove the powers of $x$ and $y$ and then evaluating the $y$ integral, which gives
\be
I(\beta,\gamma)\equiv I(z)=\frac{1}{4}\Big(1-(1+z)e^zE_1(z)\Big),
\label{eq:OTOC_2D}
\ee
with
\begin{align}\nonumber
z&\equiv \frac{i\beta\gamma}{2}=8ib_0 e^{\frac{t_2+t_4-t_1-t_3}{2}}\sinh\left(\frac{t_4-t_2}{2}\right)\sinh\left(\frac{t_3-t_1}{2}\right)\\
E_1(z)&\equiv\int_z^\infty \frac{dt}{t}e^{-t}.
\end{align}
The integral is defined by analytic continuation from $z>0$.  The result \eqref{G41} is correct in the regime
\be\label{trange1}
1 \ll t_1-t_2 <2t_{scr},
\ee
since our expression \eqref{tadv1} for the time delay is correct up to errors that are of order $\frac{\omega_2^2}{b_0^2}e^{t_1-t_2}$.  We can use time-reversal symmetry, which implies 
\be
G_4(t_1,t_2,t_3,t_4)=G(-t_1,-t_2,-t_3,-t_4)^*, 
\ee
to extend this to a formula that holds also in the opposite time regime where
\be\label{trange2}
1 \ll t_2-t_1 <2t_{scr},
\ee
giving
\be\label{G42}
G_4(t_1,t_2,t_3,t_4)\approx \frac{1}{16\pi^2}\frac{1}{\sinh^2\left(\frac{t_4-t_2}{2}\right)\sinh^2\left(\frac{t_3-t_1}{2}\right)}F(w)
\ee
with
\begin{align}
F(w)&\equiv -4w^2 I(w)=(1+w)e^{w}E_1(w)-1\nonumber\\
w&\equiv -4ib_0\frac{\sinh\left(\frac{t_4-t_2}{2}\right)\sinh\left(\frac{t_3-t_1}{2}\right)}{\sinh\left(\frac{t_2+t_4-t_1-t_3}{2}\right)}.\label{eq:omega_def}
\end{align}
\eqref{G42} is our main result for the out-of-time-order four-point function.  As a reminder it works assuming that $t_1-t_3$ and $t_2-t_4$ are order one, and also that \eqref{trange1} or \eqref{trange2} holds. If we further assume that
\be
|t_1-t_2|\ll t_{scr},
\ee
then we can simplify \eqref{G42} using the large $w$ expansion 
\be
F(w)=1-\frac{4}{w}+\ldots ,
\ee
which gives
\be
G_4(t_1,t_2,t_3,t_4)\approx \frac{1}{16\pi^2}\frac{1}{\sinh^2\left(\frac{t_4-t_2}{2}\right)\sinh^2\left(\frac{t_3-t_1}{2}\right)}\left(1-\frac{i}{b_0}\frac{\sinh\left(\frac{t_2+t_4-t_1-t_3}{2}\right)}{\sinh\left(\frac{t_4-t_2}{2}\right)\sinh\left(\frac{t_3-t_1}{2}\right)}+\ldots\right).\label{g4pres}
\ee
The leading term here is a product of two-point functions, as must be the case in free field theory, but the most important point is that the sign in front of the gravitational correction is positive, as opposed to a negative sign that was found in the black hole case \cite{Shenker:2014cwa}.  This sign arises from the negative sign of the scattering phase $\delta$, which from \eqref{advdelt} is a direct consequence of the anti-scrambling time advance.  Therefore this sign is a direct probe of anti-scrambling, so reproducing it is a key test of any proposed underlying quantum description of the de Sitter static patch.

The operator $\partial_n\chi$ we studied in this section, with matter action \eqref{eq:action_2D_matter}, has conformal dimension $1$. This is not a crucial restriction, as we could repeat the calculation using Neumann boundary conditions and $e^{i\alpha \phi}$ to get general conformal dimensions.  In fact the calculation can be done for a general matter CFT coupled to gravity, dropping any dependence on free field theory.  We explain this calculation in appendix \ref{app:OTOC_dS3}.  The result \eqref{eq:OTOC_genera_Delta} agrees with what we obtained here when the operator dimensions are taken to be $1$.

It is also interesting to consider four-point correlators where the operators are shared between two observers on the opposite Lorentzian boundaries.  We can obtain such correlators from \eqref{g4pres}, or more generally \eqref{eq:OTOC_genera_Delta}, by analytic continuation. The relevant continuation is $(t_4, t_3,t_2, t_1) = (-i\pi,-2i\epsilon+ t, -i\epsilon,t)$, which gives 
\begin{align}
    \frac{\lan M|V_RW_L(t)V_LW_L(t)|M\ran}{\lan M|V_RV_L|M\ran\lan M|W_L(t)W_L(t)|M\ran}\approx \begin{cases}
        1+\frac{1}{b_0}\frac{\Delta_V\Delta_W}{\epsilon}\cosh t & |t|\ll t_{scr}\\
        \sim \exp\Big(-\text{min}(2\Delta_W, 2\Delta_V)\,|t|\Big) & |t|\gg t_{scr}
    \end{cases}
    \label{eq:OTOC_two_sided}
\end{align}
We can consider the case where $\Delta_W\gg \Delta_V\gg 1$, in which case applying $W$ on the Hartle-Hawking state with a pair of observers creates a background on which a heavy particle is propagating. The backreacted geometry has a taller Penrose diagram. The distance between the two observers at $t=0$ initially decreases with time, which is reflected in the initial growth of \eqref{eq:OTOC_two_sided}.

This initial exponential growth is somewhat surprising, since it causes the normalized two-sided correlator to take values bigger than one. We will now argue that this cannot happen for an ordinary thermal correlator. To see this in a simple way, we will assume that $V$ and $W$ are both Hermitian and also that $V$ is unitary.   We then have
\begin{equation}
 \begin{aligned}
	0\leq & \frac{1}{Z}\Tr([e^{-\beta H/2}W(t-i\epsilon),V]^{\dagger}[e^{-\beta H/2}W(t-i\epsilon),V])\\
		=\ &2\,\Bigg[\frac{1}{Z}\Tr(e^{-\beta H}W(t-i\epsilon)W(t+i\epsilon))\frac{1}{Z}\Tr(e^{-\beta H}V^{\dagger}V)\\
        &\quad\quad\quad\quad\quad\quad\quad
	-\frac{1}{Z}\Tr\qty(e^{-\beta H}V\qty(-i\beta/2)W(t-i\epsilon)VW(t+i\epsilon))\Bigg]
\end{aligned}
\label{eq:two_sided_bound}
\end{equation}
Thus standard two-sided thermal correlators of this type are upper-bounded by the product of two-point functions, which is not true for the de Sitter correlator \eqref{eq:OTOC_two_sided}. This again suggests that if we want to obtain de Sitter correlators from a standard quantum mechanical system, we need to do something non-standard. We will come back to this issue in section \ref{sec:fold}.

So far we have discussed the four-point function in $dS_2$.  A similar calculation is possible in $dS_3$, but there is a new subtlety that arises.  This is that quantum mechanically there can be inelastic scattering out of the s-wave sector. For example two particles with angular momentum zero can scatter to a singlet state $|1,-1\ran-|00\ran+|-1,1\ran$.  This does not happen in the classical gravity scattering calculation, where spherical shells necessarily scatter to spherical shells, but the energy momentum tensor created by acting with a low-dimension operator $V(t)$ or $W(t)$ on the observer worldline is not really a classical shell: it has fluctuations that break spherical symmetry even though its expectation value is spherical.  This causes the recoil effect we discussed in section \ref{sec:recoil} to be important even for correlators that are spherically symmetric.  This effect is systematically included in \cite{Chen:2026boh,Milekhin:2026tbi}.  Our approach will instead be to simply restrict to heavy operators where the shell is sufficiently classical that the gravitational scattering dominates the calculation.  For example we can take the matter theory to be a free conformally-coupled scalar and take our boundary operators to be $\chi^n$ at large $n$.  Then the s-wave sector dominates, and as explained in section \ref{sec:solution_2D} the gravitational scattering problem is the same for $dS_2$ and $dS_3$.  Therefore our expression \eqref{eq:OTOC_genera_Delta} is correct also in $dS_3$ for operators of this type.

\section{Observer correlators II: worldline theory}\label{worldlinesec}
So far we computed only classical gravitational contributions to observer correlators.  We now introduce a more systematic formalism for JT gravity coupled to matter that allows calculations to all orders in $G$.  We have three motivations for doing so:
\bi
\item[(1)] We will see that in going to higher orders it is necessary to introduce an operational definition of proper time on the observer worldline using a physical clock, as advocated by \cite{Page:1983uc,Chandrasekaran:2022cip,Witten:2023xze}. 
\item[(2)] We will see that the calculation can be organized around the dynamics of a trio of boundary modes analogous to the single ``Schwarzian'' mode in the AdS case \cite{Jensen:2016pah,Maldacena:2016upp,Engelsoy:2016xyb}.

\item[(3)] The more precise results we obtain allow for a stronger test of the Euclidean fold prescription we introduce in the following section.
\ei
\subsection{The theory}
The 2D gravity theory we have used so far consists of a metric $g_{\mu\nu}$, a dilaton $\Phi$, and some bulk matter fields $\chi_i$.  This theory is sufficient for the tree-level two-point and four-point functions we computed, but once we include quantum corrections we need to be a bit more careful.  The first issue is that we already acknowledged the need to turn the boundary mass $M$ into a dynamical field since it changes when a particle is created or annihilated at the boundary, and it also does not need to be the same on the two sides if there is matter in between.  We will now implement this systematically by introducing a boundary field $\M(\hat{u})$, where $\hat{u}$ is a boundary time coordinate.  This however is not enough.  For one thing $\M$ needs a canonical conjugate, and we also need to deal with the fact that the boundary conditions \eqref{JTBC} preserve full boundary diffeomorphism invariance so there is no natural origin for the observer's proper time.  Both of these problems can be solved by introducing a boundary clock field $\T(\hat{u})$, conjugate to $\M$, which the observer can use to define proper time.  The action of the theory is
\be
S=\frac{1}{2}\int d^2 x \sqrt{-g}\Phi(R-2)+\int_\Gamma d \hat{u}\sqrt{-h}\big(\Phi_d-\M\big)-\int_\Gamma \T d\M +S_{matter}[\chi,g],
\ee
where $\Gamma$ is the spatial boundary (a union of two boundary worldlines) and $h$ is the induced metric on $\Gamma$. To adopt purely 2D notation we have defined
\be
\Phi_d\equiv \frac{1}{4G},
\ee
where $G$ was the three-dimensional Newton constant prior to dimensional reduction.  Varying this action with respect to $\M$ and $\T$ leads to the additional equations of motion
\begin{align}\nonumber
\dot{\M}&=0\\
\dot{\T}&=\sqrt{-h},
\end{align}
which say that the mass is constant along a worldline and that $\T$ indeed measures worldline proper time.  We will take the matter theory to be a CFT.  From now on the times of operators in observer correlators will be defined using $\T$, ensuring the full diffeomorphism-invariance of these correlators.

To compute correlation functions it is useful to introduce a Euclidean version of this theory, with action
\begin{align}
	S_E =\,&-\frac{1}{2}\int d^2 x\sqrt{g}\Phi(R-2)+\int_{\gamma_{ob}} du\sqrt{h}\qty(\mathcal{M}-\Phi_d)+\int_{\gamma_{ob}}\mathcal{\tau} d\M +S_{matter}[\chi,g]
\end{align}
where we let $\tau \equiv i\mathcal{T}$ be Euclidean clock time.  In Euclidean signature there is only a single boundary, which we call $\gamma_{ob}$.  We consider correlation functions of scalar primary operators in the matter CFT restricted to the boundary worldline.  For example in the free scalar CFT with Dirichlet boundary conditions, $\partial_n \chi$ is a primary of dimension $\Delta=1$, while with Neumann boundary conditions we can use $e^{i\alpha \chi}$ which has dimension $\Delta=\frac{\alpha^2}{4\pi}$.

As usual in JT gravity we can simplify this theory by integrating out the dilaton to impose $R = 2$. The geometry is thus a topological disk whose metric is that of a piece of a round $\mathbb{S}_2$ with metric
\be
ds^2=d\theta^2+\sin^2\theta d\phi^2.
\ee
The remaining gravitational degrees of freedom are the Euclidean clock $\tau$, the mass $\M$, and the boundary trajectory which we can parameterize as $\theta(u)$ with $u=\phi$.  We can therefore compute observer correlation functions in this theory by evaluating
\be
\lan M| \mO_n(\tau_n)\ldots \mO_1(\tau_1)|M\ran=\frac{1}{Z}\int \mathcal{D}\M\mathcal{D}\tau\mathcal{D}\theta \lan \mO_n(u(\tau_n))\ldots \mO_1(u(\tau_1)\ran_{\theta,\tau} e^{-S_{boundary}}
\ee
with
\be
S_{boundary}=\int_{\gamma_{ob}}du\sqrt{\dot\theta^2+\sin^2\theta}\qty(\mathcal{M}-\Phi_d)+\int_{\gamma_{ob}} \tau d\M
\ee
and
\be
Z=\int \mathcal{D}\M\mathcal{D}\tau\mathcal{D}\theta e^{-S_{boundary}}.
\ee
The notation $\lan \cdot \ran_{\theta,\tau}$ indicates a matter quantum field theory expectation value on a disk with a fixed boundary trajectory $\theta$ and clock function $\tau$.  The choice of boundary mass $M$ in the state $|M\ran$ enters through the integration contour for $\M$, which we take to be 
\be
\M(u)=M+\delta(u)
\ee
with $\delta(u)$ being integrated on the imaginary axis.  Since the matter correlator does not depend on $\M$ we can do the integral over $\delta$ exactly, imposing the equation of motion 
\be
\dot{\tau}=\sqrt{\dot{\theta}^2+\sin^2\theta}.
\ee
This leaves an overall shift of $\tau$ to be integrated over as the only remnant of the $\M$ and $\T$ integrals, so we can rewrite the full correlator as\footnote{The correlator in the integrand does depend on $\tau_0$ since a general trajectory is not rotationally invariant.}
\be\label{obscor}
\lan M| \mO_n(\tau_n)\ldots \mO_1(\tau_1)|M\ran=\frac{1}{\hat{Z}}\int d\tau_0\mathcal{D}\theta \lan \mO_n(u(\tau_n))\ldots \mO_1(u(\tau_1)\ran_{\theta,\tau_0} e^{-\hat{S}_{boundary}}
\ee
with
\be
\hat{S}_{boundary}=-b_0 \int_{\gamma_{ob}}du\sqrt{\dot\theta^2+\sin^2\theta},
\ee
\be
\hat{Z}=\int d\tau_0\mathcal{D}\theta \lan \mO_n(u(\tau_n))\ldots \mO_1(u(\tau_1)\ran_{\theta,\tau_0} e^{-\hat{S}_{boundary}},
\ee
and 
\be
\tau(u)=\int_0^udu'\sqrt{\dot{\theta}(u')^2+\sin^2\theta(u')}+\tau_0.\label{utau}
\ee
Here we used \eqref{b0def}.

In practical calculations the way we will use \eqref{obscor} is to write
\be
\theta=\frac{\pi}{2}-\rho
\ee
and then expand the boundary action in small $\rho$.  At quadratic order this gives
\be
\hat{S}_{boundary}=-b_0\left(1+\frac{1}{2}\int_0^{2\pi} du \left(\dot{\rho}^2-\rho^2\right)\right)+O(\rho^4).
\ee
To derive the propagator for $\rho$ it is important to note that the kinetic operator $b_0(\frac{d^2}{du^2}+1)$ has two zero modes $e^{\pm i u}$ arising from a residual sphere isometry gauge group, so removing these we have
\begin{align}\nonumber
\lan \rho(u_2)\rho(u_1)\ran&=-\frac{1}{2\pi b_0}\sum_{n\neq \pm 1}\frac{1}{n^2-1}e^{in u_{21}}\\
&=-\frac{1}{2\pi b_0}\qty(\frac{1}{2}\cos u_{21}-(\pi-u_{21})\sin u_{21})
\end{align}
with
\be
u_{21}\equiv u_2-u_1
\ee
taken to obey
\be
0< u_{21}<2\pi.
\ee
Here $\lan \cdot \ran$ means the expectation value on the right-hand side of \eqref{obscor}. 
It will also be useful to note that expanding \eqref{utau} to quadratic order and taking a difference we have
\begin{align}
      \tau_{21} =\,& u_{21}+\frac{1}{2}\int_{u_1}^{u_2}du\qty(\dot\rho^2-\rho^2)+O(\rho^4).
 \label{eq:length_renor}
\end{align}

\subsection{Two-point function}
\label{sec:two_point}
To compute the two-point observer correlator of a primary operator $\mO$ of dimension $\Delta$, from \eqref{obscor} we need to first compute the wiggly-boundary CFT correlator $\lan \mO(u(\tau_2))\mO(u(\tau_1))\ran_{\theta,\tau_0}$ and then integrate over the boundary trajectory $\rho$ and the time shift $\tau_0$.  We will first do the former using a conformal transformation that maps the wiggly boundary to a round one, and then do the integral over $\rho$ perturbatively. The $\tau_0$ integral will then be trivial.

Without gravity, the saddle location of $\gamma_{ob}$ is given by 
\begin{align}
    \theta_c(u)=\frac{\pi}{2}.
\end{align}
The correlator on this curve is given by
\begin{align}\nonumber
    \lan \mO(u(\tau_2))\mO(u(\tau_1)\ran_{\theta_c,\tau_0} =\,& \frac{C_{\Delta}}{\qty(2\sin\frac{u_{21}}{2})^{2\Delta}}\equiv K_0(u_{21}),\quad 2\pi>u_{21}>0\\
    \tau(u) =\,& u+\tau_0.\label{eq:classical}
\end{align}
Equation \eqref{eq:classical} is simply the matter correlator on the boundary of a half-sphere.

To deal with the wiggly boundary, we can do a conformal transformation $\mathcal{C}$ and map the deformed domain $\mathcal{D}_{\rho} = \Big\{\qty(u, \theta):\theta<\frac{\pi}{2}-\rho(u)\Big\}$ back to $\mathcal{D}_0$. Restricting to the boundary, we have a reparametrization
\begin{align}
	\mathcal{C}\big|_{\gamma_{ob}}:\ \qty(u,\frac{\pi}{2}-\rho(u)) \longmapsto \qty(u+\xi(u), \frac{\pi}{2}).
\end{align}
The details are in appendix \ref{app:conformal}, with the relation between $\rho$ and $\xi$ being given by \eqref{xirho}. The propagator of $\xi$ is given by
\begin{align}
\lan \xi(u_2)\xi(u_1)\ran&=-\frac{1}{2\pi b_0}\sum_{n\neq 0,\pm 1}\frac{1}{n^2-1}e^{in u_{21}}\nonumber\\
&=-\frac{1}{2\pi b_0}\qty(1+\frac{1}{2}\cos u_{21}-(\pi-u_{21})\sin u_{21})\label{eq:prop_xi}
\end{align}
for $2\pi>u_{21}>0$.

For an operator $\mathcal{O}$ with conformal dimension $\Delta$, the correlation function on the deformed boundary is given by
\begin{align}\nonumber
    \lan \mO(u_2)\mO(u_1)\ran_{\theta}=\,& \qty(\frac{1+\dot\xi(u_2)}{\sqrt{h(u_2)}})^{\Delta}\qty(\frac{1+\dot\xi(u_1)}{\sqrt{h(u_1)}})^{\Delta}K_{0}(u_{21}+\xi(u_2)-\xi(u_1))\\
	=\,&K_{0}(u_{21})\qty[1+\Delta\,\mathcal{B}^{(1)}(u_1, u_2)+\Delta\,\mathcal{B}^{(2)}(u_2,u_1)+\frac{\Delta^2}{2}\mathcal{B}^{(1)}(u_2, u_1)^2+O(\rho^3)]\label{Otheta}
\end{align}
where 
\begin{align}\nonumber
    \mathcal{B}^{(1)}(u_2, u_1) =\,& \dot \xi_2+\dot \xi_1-(\xi_2-\xi_1)\cot(\frac{u_{21}}{2})\\
    \mathcal{B}^{(2)}(u_2, u_1)=\,&\frac{1}{4}\qty(-2\qty(\dot\xi_2^2+\dot \xi_1^2)+\frac{(\xi_2-\xi_1)^2}{\sin^2(\frac{u_{21}}{2})})-\frac{1}{2}\qty(\dot\rho_1^2-\rho_1^2+\dot\rho_2^2-\rho_2^2)\label{eq:B2}\\\nonumber
     \mathcal{B}^{(1)}(u_2, u_1)^2=\,&\dot \xi_2^2+\dot \xi_1^2
  +2\dot\xi_2\dot\xi_1+(\xi_2^2+\xi_1^2-2\xi_2\xi_1)\cot^2\qty(\frac{u_{21}}{2})\\
  &\quad-2\qty(\dot \xi_2\xi_2-\dot \xi_2\xi_1+\dot \xi_1\xi_2-\dot \xi_1\xi_1)\cot(\frac{u_{21}}{2}).\label{eq:B12}
\end{align}
Clearly $\expval{\mathcal{B}^{(1)}(u_2, u_1)}=0$ due to the vanishing of the one-point function of $\xi$. On the other hand, from equations \eqref{eq:length_renor}, \eqref{eq:B2}, and \eqref{eq:B12} we see that $\lan \tau(u)\ran$, $\lan \mathcal{B}^{(2)}(u_2, u_1)\ran $, and $\lan \mathcal{B}^{(1)}(u_2, u_1)^2\ran$ will have UV divergences due to product operators evaluated at coincident points so some renormalization is necessary when taking expectation values. 

Now say that we introduce a renormalization scheme where we take $\expval{\dot\rho^2}_{\mathrm{ren}} =\expval{\dot\xi^2}_{\mathrm{ren}} = -\frac{q}{4\pi b_0}$.\footnote{We use the same renormalization scheme because these are related by a Hilbert transform. See \eqref{eq:xi_rho}.} From \eqref{eq:prop_xi}, we then have
\begin{align}\nonumber
    \expval{\mathcal{B}^{(1)}(u_2, u_1)^2}=\,&-\frac{q+3}{2\pi b_0}\\
    \expval{\mathcal{B}^{(2)}(u_2, u_1)}=\,&\frac{q-1}{2\pi b_0}-\frac{\pi-u_{21}}{2\pi b_0}\tan(\frac{\pi-u_{21}}{2}),\label{B12}
\end{align}
and using 
\begin{align}
     \expval{\tau_{21}} =\,& u_{21}-\frac{u_{21}}{2}\frac{q-1}{4\pi b_0}
\end{align}
we also have
\begin{align}
    \lan K_{0}(u(\tau_2)-u(\tau_1))\ran\approx\,&
  K_{0}(\tau_{21})\qty[1-\frac{\Delta(q-1)}{8\pi b_0}\tau_{21}\cot(\frac{\tau_{21}}{2})]+O\qty(\frac{1}{b_0^2})\label{K0u}
\end{align}
The terms that depend on $q$ here are regularization-scheme dependent.  We will choose a scheme so that $\tau_{\mathrm{ren}}$ has period $2\pi$. At one-loop order this amounts to setting $q=1$.\footnote{At higher loop order the value of $q$ will be further adjusted.}  Combining \eqref{B12} and \eqref{K0u} we can compute the expectation value of \eqref{Otheta}, giving the leading correction to the two-point function as
\begin{align}\nonumber
   &\lan M|\mathcal{O}(\tau_2)\mathcal{O}(\tau_1)|M\ran\\
   =\,&\frac{C_{\Delta}}{\qty(2\sin\qty(\frac{\tau_{21}}{2}))^{2\Delta}}\qty[1-\frac{\Delta}{2\pi b_0}\qty(\pi-\tau_{21})\cot(\frac{\tau_{21}}{2})-\frac{\Delta^2}{\pi b_0}+O(1/b_0^2)]\label{eq:correct_two_point_Euclidean_0}
\end{align}
In terms of a Lorentzian clock, we have
\begin{align}
    &\lan M|\mathcal{O}(\mathcal{T}_2)\mathcal{O}(\mathcal{T}_1)|M\ran \nonumber\\
=\,&\frac{C_{\Delta}}{\qty(2i\sinh\qty(\frac{\mathcal{T}_{21}}{2}-i\epsilon))^{2\Delta}}\qty[1+\frac{\Delta}{2\pi b_0}\qty(\frac{\mathcal{T}_{21}+i\pi}{\tanh(\frac{\mathcal{T}_{21}}{2})})-\frac{\Delta^2}{\pi b_0}+O(1/b_0^2)]\label{eq:correct_two_point_Lorentzian}
\end{align}

\subsection{Four-point function}
We now revisit the leading-order gravitational corrections to the connected piece of the four-point function using this boundary action approach. We will consider the four-point function of two $V$ operators and two $W$ operators, both of which are conformal scalar primaries.  To avoid other interactions we will assume the theory is free, so that the matter correlators factorize into products of two-point functions.  As there are no coincident points at this order, no regularization is needed.  We will begin with the correlator $\lan M|V(\tau_4)V(\tau_3)W(\tau_2)W(\tau_1)|M\ran$, with $2\pi>\tau_4>\tau_3>\tau_2>\tau_1>0$.  The only connected contribution to the four-point function comes from
\begin{align}
    \expval{\mathcal{B}^{(1)}(\tau_4, \tau_3)\mathcal{B}^{(1)}(\tau_2, \tau_1)}=0,
\end{align}
so we have
\begin{align}\nonumber
&\lan M| V(\tau_4)V(\tau_3)W(\tau_2)W(\tau_1)|M\ran\\
=\,&\lan M|V(\tau_4)V(\tau_2)|M\ran\lan M|W(\tau_3)W(\tau_1)|M\ran+O\qty(\frac{1}{b_0^2})  
\end{align}
Thus at order $\frac{1}{b_0}$, the connected part of this four-point function vanishes. 

Next we consider the alternating order  $\expval{V(\tau_4)W(\tau_3)V(\tau_2)W(\tau_1)}$,again with $2\pi>\tau_4>\tau_3>\tau_2>\tau_1>0$.  This is the correlator that analytically continues to the out-of-time-order correlator we studied in the previous section.  Its connected part is controlled by
\begin{align}
     \expval{\mathcal{B}^{(1)}(\tau_4, \tau_2)\mathcal{B}^{(1)}(\tau_3, \tau_1)}=\frac{1}{b_0}\frac{\sin(\frac{\tau_4+\tau_2}{2}-\frac{\tau_3+\tau_1}{2})}{\sin(\frac{\tau_4-\tau_2}{2})\sin(\frac{\tau_3-\tau_1}{2})},
\end{align}
As a result 
\begin{align}\nonumber
   & \lan M|V(\tau_4)W(\tau_3)V(\tau_2)W(\tau_1)|M\ran\\\nonumber
=\,&\lan M|V(\tau_4)V(\tau_2)|M\ran\lan M|W(\tau_3)W(\tau_1)|M\ran\times\\
&\quad\quad\qty(1+\frac{\Delta_V\Delta_W}{b_0}\frac{\sin(\frac{\tau_4+\tau_2}{2}-\frac{\tau_3+\tau_1}{2})}{\sin(\frac{\tau_4-\tau_2}{2})\sin(\frac{\tau_3-\tau_1}{2})})+\mathcal{O}\qty(\frac{1}{b_0^2}).  
\end{align}

Going back to Lorentzian clock time, we have
\begin{align}\nonumber
   & \lan M|V(\T_4)W(\T_3)V(\T_2)W(\T_1)|M\ran\\\nonumber
=\,&\lan M|V(\T_4)V(\T_2)|M\ran\lan M|W(\T_3)W(\T_1)|M\ran\times\\
&\quad\quad\qty(1-i\frac{\Delta_V\Delta_W}{b_0}\frac{\sinh(\frac{\T_4+\T_2}{2}-\frac{\T_3+\T_1}{2})}{\sinh(\frac{\T_4-\T_2}{2})\sinh(\frac{\T_3-\T_1}{2})})+\mathcal{O}\qty(\frac{1}{b_0^2}) . 
\end{align}
Reassuringly this agrees with \eqref{g4pres} with $\Delta=1$, but we have also learned something new: apparently the only corrections to \eqref{g4pres} for times that are of $O(1)$ (recall that there we assumed $|t_2-t_1|\gg 1$) are the corrections \eqref{eq:correct_two_point_Lorentzian} to the two-point functions multiplying the correlator.

\section{Anti-scrambling from Euclidean folds}  
\label{sec:fold}
Let's now gather our gravitational results for two-point and four-point correlation functions as a function of Euclidean clock time $\tau$: 
\be
G_2(\tau_{21})=\frac{C_\Delta}{\left(2\sin\frac{\tau_{21}}{2}\right)^{2\Delta}}\left(1-\frac{\Delta}{2\pi b_0}\qty(\pi-\tau_{21})\cot(\frac{\tau_{21}}{2})-\frac{\Delta^2}{\pi b_0}+\ldots\right)\label{G2euc}
\ee
for the two-point function of a bulk operator of conformal dimension $\Delta$ and 
\be
G_4(\tau_1,\tau_2,\tau_3,\tau_4)=\frac{1}{16\pi^2}\frac{1}{\sin^2 \left(\frac{\tau_{42}}{2}\right)\sin^2\left(\frac{\tau_{31}}{2}\right)}F(w)\label{G4euc}
\ee
with
\be\label{wfold}
w=-4b_0\frac{\sin \left(\frac{\tau_{42}}{2}\right)\sin\left(\frac{\tau_{31}}{2}\right)}{\sin\left(\frac{\tau_{43}+\tau_{21}}{2}\right)}
\ee
for the four-point function of $\partial_n\chi_i$ with the flavors chosen as in \eqref{G4def}.  If all $\tau$ are real, or more generally when $|t_2-t_1|\ll t_{scr}$, we have the large-$w$ expansion
\be
F(w)\approx \left(1+\frac{1}{b_0}\frac{\sin\left(\frac{\tau_{43}+\tau_{21}}{2}\right)}{\sin\left(\frac{\tau_{42}}{2}\right)\sin\left(\frac{\tau_{31}}{2}\right)}+\ldots\right).\label{FwE}
\ee
The two-point function \eqref{G2euc} as far as we can tell is consistent with being the two-point function of a Heisenberg operator in some hypothetical underlying quantum description of the de Sitter static patch.  It is positive, analytic in $\tau_1$ and $\tau_2$, and obeys the KMS condition with periodicity $2\pi$.  The four-point function however exhibits anti-scrambling: in the expansion \eqref{FwE} the sign of the first correction is opposite to the sign found in the black hole case, and indeed more generally \eqref{G4euc} is a perfect match for the out-of-time-order correlation function computed for black holes in the eikonal scattering regime except for the replacement $F(w)\to F(-w)$, see equation 6.57 of \cite{Maldacena:2016upp} (setting $\Delta=1$ and $C=b_0$). 

What kind of calculation in an underlying quantum theory could give rise to these correlation functions?  We need to do something unusual that somehow gives a normal two-point function but an  anti-scrambling four-point function.  We do not have a complete proposal for how to do this, but as we now explain one possibility is to use correlation functions that are folded in Euclidean signature.  There is a long history of computing correlation functions in Lorentzian signature with folded time, meaning operators that are not multiplied in time order, but in Euclidean signature there is a standard reason not to do this: in a quantum system with an energy that is not bounded from above the operator $e^{\tau H}$  with $\tau>0$ is quite badly-behaved in the sense that its domain does not include the states that are obtained by acting on the ground state with a finite number of Heisenberg operators.  To avoid this problem we will simply need to postulate that we are considering a quantum system whose Hamiltonian is bounded from both below and above.  This is related to the question of which quantum systems make sense at negative temperature, which is a point we will return to at the end of the section.

\begin{figure}
    \centering
    \includegraphics[width=0.8\linewidth]{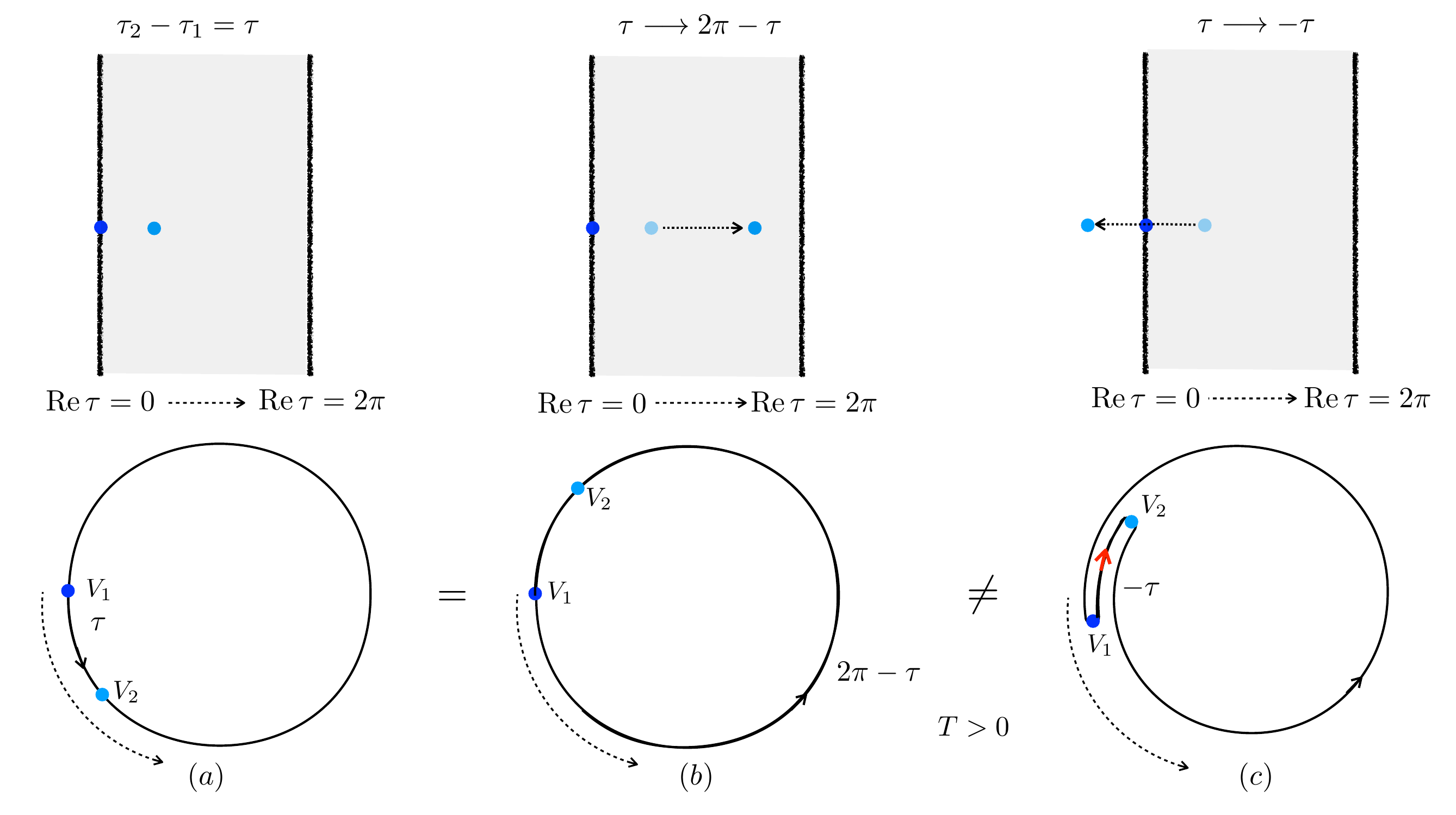}
    \caption{Reversing the Euclidean time-separation of a two-point function in two different ways: going ``around the back'' of the thermal circle, which leads to the same value for the correlator, or analytically continuing across the operator using a Euclidean fold.  The latter requires a finite system due to the backwards Euclidean evolution indicated with a red arrow, and gives a different value even though the two operators appear with the same separation in case (b) and case (c).}
    \label{fig:fold_two_point_function}
\end{figure}
We can motivate our rule by the fact we just mentioned that the scrambling black hole four-point function and the anti-scrambling de Sitter four-point function (after dividing by a product of two-point functions in each case) are related by the replacement $w\to -w$.  Looking at \eqref{wfold}, we can bring about this replacement by introducing an operation that reverses the sign of $\sin\left(\frac{\tau_{42}+\tau_{31}}{2}\right)$ but not the sign of $\sin\left(\frac{\tau_{42}}{2}\right)\sin\left(\frac{\tau_{31}}{2}\right)$.  More concretely starting with a four-point function in an ordinary quantum system, we can get an anti-scrambling four-point function by analytically continuing $\tau_1^{QM},\ldots \tau_4^{QM}$ in a way that brings about these sign changes.  To find an analytic continuation that works, it is useful to first consider the domain of analyticity of the two-point function \eqref{G2euc}.  Viewing it as a function purely of $\tau_{21}$, there is a singularity at $\tau_{21}=0$ when the operators come together but it is analytic in the strip
\be
0<\mathrm{Re}\,\tau_{21}<\beta
\ee
with $\beta=2\pi$.  For any thermal correlator 
\be
\Tr\left(e^{-\beta H} O_n(\tau_n)\ldots O_1(\tau_1)\right)
\ee
of a quantum system computed in the standard Euclidean order
\be
\tau_1<\tau_2\ldots <\tau_n,
\ee 
the operator $e^{-\tau H}$ appears only with $\tau$ in this strip.  Moreover within this strip $G_2$ obeys the condition
\be\label{kms}
G_2(2\pi-\tau_{21})=G_2(\tau_{21}),
\ee
which is justified by moving the operator at $\tau_2$ ``the back way around the thermal circle'' without crossing the operator at $\tau_1$ as shown in figure \ref{fig:fold_two_point_function}. This equation is a special case of the KMS condition.  It is to be distinguished from what we would get by moving $\tau_2$ across the operator at $\tau_1$, which in a finite system where $H$ is bounded from above and below leads to a Euclidean fold as shown in figure \ref{fig:fold_two_point_function}.  In the two-point function there is no need to actually introduce such a fold, but we will now see that in the four-point function similar folds play a crucial role.  

\begin{figure}
    \centering
    \includegraphics[width=0.8\linewidth]{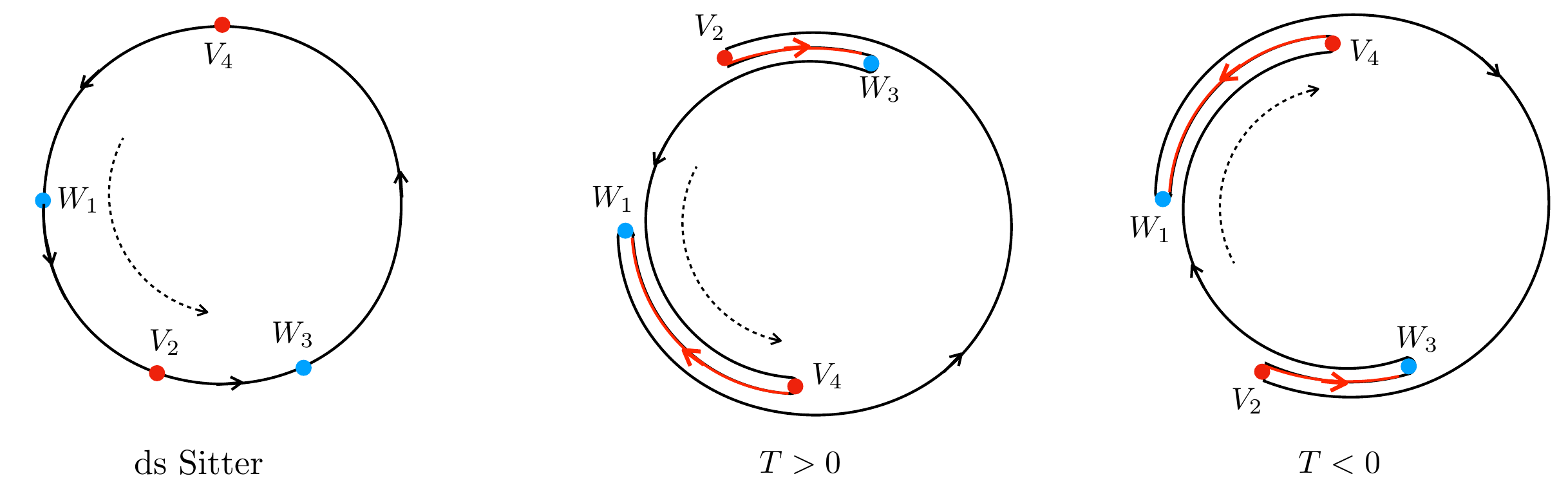}
    \caption{Euclidean folds and the de Sitter four-point function.  On the left is a Euclidean dS observer correlator in standard Euclidean order.  In the middle and on the right are folded correlators in a standard quantum system, using the $T>0$ prescription \eqref{QM2dS} and the $T<0$ prescription \eqref{QM2dS2}}
    \label{fig:fold_OTOC_0}
\end{figure}
Turning now to the four-point function, our proposal is that a de Sitter-like anti-scrambling four-point function can be computed by starting with an ordinary thermal four-point function in a many-body quantum system of bounded energy by doing the continuation
\be\label{QM2dS}
(\tau^{QM}_1,\tau^{QM}_2,\tau^{QM}_3,\tau^{QM}_4)=(-\tau_1^{dS},2\pi n-\tau_2^{dS},2\pi-\tau_3^{dS},2\pi(n+1)-\tau_4^{dS})
\ee
where $n$ is an arbitrary integer.  This leads to a folded Euclidean correlation function, shown for $n=1$ in the middle of figure \ref{fig:fold_OTOC_0}.  The reason this continuation works is that it implies
\begin{align}\nonumber
\tau_{31}^{QM}&=2\pi-\tau_{31}^{dS}\\\nonumber
\tau_{42}^{QM}&=2\pi-\tau_{42}^{dS}\\
\tau_{43}^{QM}+\tau_{21}^{QM}&=4\pi n-\tau_{43}^{QM}-\tau_{21}^{QM}.
\end{align}
These replacements do not change the product of two-point functions appearing in front of the four-point function due to \eqref{kms}, but they do change the sign of $w$ and thus convert a scrambling four-point function to an anti-scrambling one.  At the moment we do not have a way of selecting a particular value of $n$, which is representative of the broader point that we do not at the moment know how to turn this observation into a general rule for computing all de Sitter correlators from a standard quantum system.  For example the prescription \eqref{QM2dS} is tied to the specific ordering of the operators in the de Sitter correlator we are trying to compute.  We also note that in order for this continuation to really do its job, it must be the case that the analytic continuation of the semiclassical scrambling answer to the antiscrambling one indeed commutes with the semiclassical limit of the hypothetical underlying quantum system.  This assumption could fail if there is some kind of Stokes' phenomenon that disrupts the analytic continuation to our semiclassical expressions \eqref{G2euc} and \eqref{G4euc}.  It would be interesting to study the validity of this assumption in a concrete model such as the SYK model.

In section \ref{sec:OTOC_1} we derived a bound for a standard two-sided correlator in a quantum mechanical system. That bound is violated by de Sitter correlators. The reason is that with a negative Euclidean fold present, the de Sitter two-sided correlator no longer computes the connected term in the commutator square given in \eqref{eq:two_sided_bound}. 

One aspect of our Euclidean fold prescription \eqref{QM2dS} that is somewhat mysterious is that it reverses the times of the operators.  We can actually dispense with this by making the replacements
\begin{align}\nonumber
\tau^{QM}&\to -\tau^{QM}\\\nonumber
\beta&\to -\beta\\
H&\to -H
\end{align}
in the hypothetical underlying system, since this operation leaves all correlation functions invariant.  Usually we do not consider this transformation because it does not act within the set of Hamiltonians that are bounded from below, but for a finite system both descriptions are equally natural.  We can therefore use the perhaps more intuitive analytic continuation
\be\label{QM2dS2}
(\tau^{QM}_1,\tau^{QM}_2,\tau^{QM}_3,\tau^{QM}_4)'=(\tau_1^{dS},\tau_2^{dS}-2\pi n,\tau_3^{dS}-2\pi,\tau_4^{dS}-2\pi(n+1)),
\ee
at the cost of using an ensemble with negative temperature.  We show how the folds work after this replacement in figure \ref{fig:fold_OTOC_0}.  At present we do not have a way of deciding which of these prescriptions (if any) is correct, but we note that there is a natural sense in which the temperature of the de Sitter static patch is negative: increasing the energy of the matter in the static patch decreases its entropy due to the horizon decrease in \eqref{coshor}.   

Perhaps the most important defect of this proposal is that it is only able to account for the milder form of anti-scrambling that happens in the regime where $|t_1-t_2|<2t_{scr}$.  We discussed a more extreme form of anti-scrambling in section \ref{2shellsec}: for time separations that are larger than $2t_{scr}$ the back-reaction from the shockwave scattering is sufficient to destroy the entire dS universe.  This is an even more striking feature that any underlying quantum system must reproduce, and we hope to return to this problem in the future.  Another thing to consider is to what extent the recoil effect can or should be reproduced by this underlying quantum description.

\section*{Acknowledgement}
We thank Netta Engelhardt, David Kolchmeyer, Henry Lin, Juan Maldacena, Leonard Susskind, and Misha Usatyuk for discussions. 
DH is supported by the Packard Foundation as a Packard Fellow, the US
Department of Energy under grants DE-SC0012567 and DE-SC0025937, and the MIT
department of physics. YZ is supported by the John Templeton Foundation Award 41001491-013 (subagreement under prime agreement No. ID \# 63670), DOE grant DE-SC0012567,  and the ``Algebras and Complexity in Quantum Gravity and Field Theory'' award \#DE-SC0025937.

\appendix
\section{JT Calculations}\label{JTapp}
In this appendix we give some details for the derivations of section \ref{sec:solution_2D}. In the Kruskal coordinates \eqref{kruskal}, the dilaton equations of motion are
\begin{align}\nonumber
\partial_\mp^2\left((1-X^+X^-)\Phi\right)&=-(1-X^+X^-)T_{\mp\mp}\\
-\partial_+\partial_-\Phi+\frac{2}{(1-X^+X^-)^2}\Phi&=-T_{+-},\label{dilatonkruskal}
\end{align}
from which it is not difficult to confirm the solution \eqref{dilsol}.  The jump conditions \eqref{Dabc} are obtained by integrating these equations of motion across the shell energy-momentum tensor \eqref{Tnull}.  Considering a right/left moving shell of strength $k$ at position $X_0^{\mp}$, integrating the first line of \eqref{dilatonkruskal} over $X^{\mp}$ using \eqref{dilsol} gives
\be
\Delta a_\mp+\Delta b X^{\pm} =-k(1-X^\pm X_0^{\mp}),
\ee
while the continuity of $(1-X^+X^-)\Phi$ gives
\be
\Delta a_\pm X^\pm + \Delta a_{\mp}X_0^\mp+\Delta b(1+X^\pm X_0^\mp).
\ee
Solving these equations for arbitrary $X^\pm$ gives \eqref{Dabc}.

The energy \eqref{shellE} of a shell in the rest frame of the boundary is derived from
\be
\omega=\int dX^{\mp}T_{\mp\mp}\xi^{\mp},
\ee
where
\be
\xi=\dot{X}^+\partial_++\dot{X}^-\partial_-
\ee
is the Killing vector that reduces to proper time evolution along the boundary.  

The boundary trajectory is determined by solving 
\be
a_+X^++a_-X^-+b(1+X^+X^-)=0,
\ee
and solving this together with
\be
\frac{2\dot{X}^+\dot{X}^-}{(1-X^+X^-)^2}=1
\ee
determines $X^\pm(t)$.

\section{Four-point function with general scaling dimensions}
\label{app:OTOC_dS3}
\begin{figure}
    \centering
    \includegraphics[width=0.5\linewidth]{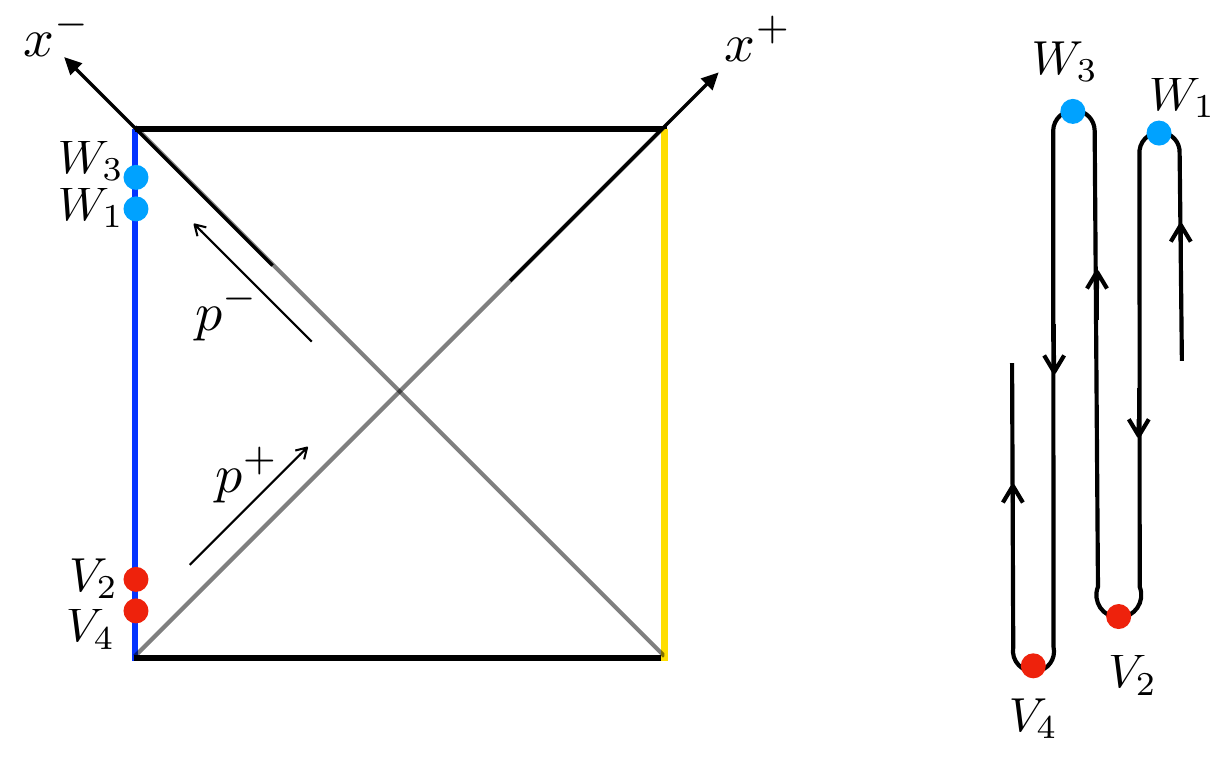}
    \caption{A general one-sided out-of-time order four-point function.}
    \label{fig:OTOC_bulk}
\end{figure}
In this appendix we generalize our discussion of the $dS_2$ out-of-time-order four-point function in section \ref{sec:OTOC_1} to operators with general boundary scaling dimension, 
\begin{align*}
	G(t_1,t_2,t_3,t_4) = \lan M|V(t_4)W(t_3)V(t_2)W(t_1)|M\ran,
\end{align*}
with scaling dimensions $\Delta_V$ and $\Delta_W$.  See figure \ref{fig:OTOC_bulk} for an illustration.  The idea is to write
\begin{align}\nonumber
W(t)|M\ran&=\int_{-\infty}^0 \frac{dp_+}{2\pi} \psi_W(p_+,t)|p_+,W\ran\\
V(t)|M\ran&=\int_{-\infty}^0 \frac{dp_-}{2\pi}\psi_V(p_-,t)|p_-,V\ran
\end{align}
where $|p_+,W\ran$ and $|p_-,V\ran$ are simultaneous eigenstates of $P_\pm$  respectively, and which we give the Lorentz-invariant normalizations
\begin{align}\nonumber
\lan p_+',W|p_+,W\ran&=-4\pi p_+\delta(p_+-p_+')\\ 
\lan p_-',V|p_-,V\ran&=-4\pi p_-\delta(p_--p_-').\label{boostip}
\end{align}
Using these expressions we then have the leading gravitational contribution to the four-point function as
\be
G(t_1,t_2,t_3,t_4)=4\int_{-\infty}^0\frac{dp_-}{2\pi}\int_{-\infty}^0\frac{dp_+}{2\pi}p_+p_-\psi_W(p_+,t_1)\psi_W(p_+,t_3)^*\psi_V(p_-,t_2)\psi_V(p_-,t_4)^* e^{i\delta(p_+p_-)},
\ee
where the scattering phase is given by \eqref{eq:phase} as before.  To compute the four-point function we therefore need only compute the wave functions $\psi_W$ and $\psi_V$.  In the main text we did this by expanding the field in Klein-Gordon modes, but for a general CFT this method is not available.  As explained in \cite{Shenker:2014cwa} however, they can also be extracted from a bulk-boundary two-point function with the bulk point evaluated on the horizon.  The key assumption is that there are bulk scalar primary operators whose two-point functions with  $V$ and $W$ are not zero (otherwise we do not get a signal that propagates into the bulk).  For convenience we will also assume the bulk operator has the same scaling dimension as the boundary operator, although this is not necessary.  In the latter case the bulk-boundary two-point function is determined by bulk conformal symmetry to be
\begin{align}
    \lan M|\mO_{bulk}(X)\mO_{boundary}(t)|M\ran= \frac{C_{\Delta}}{\qty[2\qty(1-\frac{e^{-(t+i\epsilon)}X^--e^{t+i\epsilon}X^+}{1-X^+X^-})]^{\Delta}}.
\end{align}
To extract the wave functions we compute
\begin{align}\nonumber
\int &dX^\pm e^{-ip_\pm X^\pm}\lan M|\mO_{bulk}(X)\mO_{boundary}(t)|M\ran|_{X^\mp=0}\\\nonumber
&=\int dX^\pm e^{-ip_\pm X^\pm}\int \frac{dp_\pm'}{2\pi}\psi_{\mO_{boundary}}(p_\pm ',t)\lan M|\mO_{bulk}(0)|p_\pm,\mO_{boundary}\ran e^{ip_\pm' X^\pm}\\
&=\psi_{\mO_{boundary}}(p_\pm,t)\lan M|\mO_{bulk}(0)|p_\pm,\mO_{bulk}\ran.
\end{align}
Finally we note that by boost invariance the matrix element $\lan M|\mO_{bulk}(0)|p_\pm,\mO_{bulk}\ran$ is actually independent of $p_\pm$ (here it is important that we used the boost-invariant norm \eqref{boostip}).  Therefore we have
\begin{align}
\psi_W(p^{-},t) \propto\,& \int dx^+ e^{2ip^- x^+}\qty[2\qty(1+ e^{t_1+i\epsilon}x^+)]^{-\Delta_W}\nonumber\\
=\,& \frac{\pi}{\Gamma(\Delta_W)}e^{\frac{i\pi\Delta_W}{2}}e^{-\Delta_W t_1}\exp(-2i p^- e^{-t_1})\theta(p^-)(p^-)^{\Delta_W-1}\\
	\psi_V(p^{+},t) \propto\,& \int dx^- e^{2ip^+ x^-}\qty[2\qty(1- e^{-(t_2+i\epsilon)}x^-)]^{-\Delta_V}\nonumber\\
    =\,& \frac{\pi}{\Gamma(\Delta_V)}e^{-\frac{i\pi\Delta_V}{2}}e^{\Delta_V t_2}\exp(2i p^+ e^{t_2})\theta(p^+)(p^+)^{\Delta_V-1}.
\end{align}
From here the calculation is identical to the calculation in section \ref{sec:OTOC_1}, replacing \eqref{eq:I} by
\be
I_{\Delta_V,\Delta_W}(\beta,\gamma)\equiv\int_0^\infty dx \int_0^\infty dy \,x^{2\Delta_V-1}y^{2\Delta_W-1}\,e^{-i\beta x-i\gamma y-2ix y}.
\label{eq:I_2}
\ee
The normalized four-point function is given by 
\begin{align}
	\frac{\expval{V(t_4)W(t_3)V(t_2)W(t_1)}}{\expval{V(t_4)V(t_2)}\expval{W(t_3)W(t_1)}}=w^{2\Delta_W}U\qty(2\Delta_W, 2\Delta_W+1-2\Delta_V, w)
    \label{eq:OTOC_genera_Delta}
\end{align}
where $w$ was given in \eqref{eq:omega_def} with $b_0 = \frac{S}{2\pi}$ and $U$ is the confluent hypergeometric function
\be
U(a,b,w)=\frac{1}{\Gamma(a)}\int_0^\infty ds \,e^{-ws}s^{a-1}(1+s)^{b-a-1}.
\ee
The large $w$ behavior is given by
\be
\frac{\expval{V(t_4)W(t_3)V(t_2)W(t_1)}}{\expval{V(t_4)V(t_2)}\expval{W(t_3)W(t_1)}}=\left(1-\frac{4\Delta_V\Delta_W}{w}+\ldots\right),\label{eq:OTOC_short}
\ee
again with $w$ defined by \eqref{eq:omega_def}.  As discussed in section \ref{sec:OTOC_1}, the negative sign in \eqref{eq:OTOC_short} shows anti-scrambling.

\section{Conformal transformation of correlators}

\label{app:conformal}
We write the round sphere metric as
\begin{align}
    ds^2 = d\theta^2+\sin^2\theta d\phi^2
\end{align}

Consider the domain
\begin{align}
	\mathcal{D}_{\rho} = \Big\{\qty(\phi, \theta):\theta<\frac{\pi}{2}-\rho(\phi)\Big\}
\end{align}
and the conformal map 
\begin{align}
	\mathcal{C}:\mathcal{D}_\rho\longrightarrow \mathcal{D}_0
\end{align}
with boundary restriction
\begin{align}\nonumber
\mathcal{C}\big|_{\gamma_{ob}}:\gamma_{ob}&\longrightarrow \gamma_0\\
\qty(u,\frac{\pi}{2}-\rho(u)) &\longmapsto\qty(f_{\rho}(u), \frac{\pi}{2}).
\end{align}

The induced metric on $\gamma_{ob}$ is given by
\begin{align}
d\tilde s_{\gamma_{ob}}^2=f_{\rho}'(u)^2 du^2=\Omega_{\partial}^2 \,ds_{\gamma_{ob}}^2
\end{align}
where
\begin{align}
    \Omega_{\partial}(u) = \frac{|f_{\rho}'(u)|}{\sqrt{h(u)}} =\frac{|f_{\rho}'(u)|}{\sqrt{\dot\rho^2+\cos^2\rho}}
\end{align}

For an operator $\mathcal{O}$ with conformal dimension $\Delta$, its correlator is invariant under $\mathcal{O}(u)\big|_{ds_{\gamma_{ob}}}\longrightarrow \mathcal{\tilde O}(u)\big|_{d\tilde s_{\gamma_{ob}}}\equiv \Omega_{\partial}^{\Delta}\mathcal{O}(f(u))\big|_{ds_{\gamma_0}}$.
Hence we have
\begin{align}\nonumber
    \lan \mO(u_2)\mO(u_1)\ran_{\theta}  =\,& \Omega_{\partial}(u_2)^{\Delta}\Omega_{\partial}(u_1)^{\Delta}K_{0}(f(u_2)- f(u_1))\\
    =\,&\qty(\frac{f_{\rho}'(u_2)}{\sqrt{h(u_2)}})^{\Delta}\qty(\frac{f_{\rho}'(u_2)}{\sqrt{h(u_1)}})^{\Delta}K_{0}\qty(f_{\rho}(u_2)-f_{\rho}(u_1))
\end{align}

Next, we work out $\xi(u)\equiv f_{\rho}(u)-u$ in terms of $\rho(u)$. 

\subsection{Conformal map}
We work in conformal coordinates. Let $\tanh y = \cos\theta$. The round sphere metric becomes
\begin{align}
   ds^2 = \frac{1}{\cosh^2 y}\qty(dy^2+du^2), 
\end{align}
and the regions are given by
\begin{align}\nonumber
    \mathcal{D}_0 =\,& \big\{(u,y): y>0\}\\
    \mathcal{D}_{\rho} =\,&\big\{(u,y): y>\arctanh(\sin\rho(u))\equiv y_\rho(u)\big\}
\end{align}
Let $z= u+iy$, $\bar z = u-iy$. We look for a conformal transformation $\mathcal{C}(z)$ such that 
\begin{align}
    \Im \mathcal{C}(u+i y_{\rho}(u))=0.
\end{align}
We work to linear order in $\rho(u)$. Let $\mathcal{C}(z) = z+\zeta(z)$.
\begin{align}\nonumber
   & y_{\rho}(u)+\Im \zeta(u+iy_{\rho}(u)) = 0\\
   &\Im \zeta(u) = -\rho(u)+\mathcal{O}(\rho^2)
\end{align}
So we have
\begin{align}
    f(u) =\,& \Re\mathcal{C}(u+i y_{\rho}(u)) = u+ \Re\zeta(u+iy_{\rho}(u)) = u+\Re \zeta(u)
\end{align}
To linear order in $\rho$, $\xi(u) = \Re\zeta(u)$. As $\zeta$ is a holomorphic function, up to a constant, $\xi$ is the Hilbert transform of $\rho$:
\begin{align}\label{xirho}
    \xi(u) = \frac{1}{2\pi}\mathrm{PV}\int_0^{2\pi}du'\frac{\rho(u')}{\tan(\frac{u-u'}{2})}
\end{align}
To see this, note that in terms of Fourier modes $ \zeta(z) = \frac{1}{2\pi}\sum_{n} \zeta_n e^{inz}$, we have $\xi_n=0$ for $n<0$, which follows from analyticity of the function $\zeta(z)$ in the upper half plane.

On the other hand, from
\begin{align}
    \zeta(u) = \xi(u)-i\rho(u),\quad \zeta_n = \xi_n-i\rho_n
\end{align}
we have $\xi_n = i\rho_n$ for $n<0$. As $\xi(u)$ and $\rho(u)$ are both real functions, we also have
\begin{align}
    \xi_n = \xi_{-n}^*,\quad \rho_n = \rho_{-n}^*
\end{align}
As a result, for $n>0$, we have
\begin{align}
    \xi_n = \xi_{-n}^* = \qty(i\rho_{-n})^* = -i\rho_n
\end{align}
Putting everything together, we have\footnote{We require $\xi(u)=0$ when $\rho(u)=0$. This implies $\xi_0=0$.}
\begin{align}\nonumber
    \xi_n =\,& -i\,\mathrm{sign}(n)\rho_n,\quad n\neq 0\\
    \xi_0=\,&0 \label{eq:xi_rho}
\end{align}
In coordinate space, we have
\begin{align}\nonumber
    \xi(u)=\,& -\frac{i}{2\pi}\sum_{n>0}\qty(\rho_n e^{inu}-\rho_{-n}e^{-inu})\\\nonumber
    =\,&-\frac{i}{2\pi}\sum_{n>0}\qty(\int_0^{2\pi}du' \rho(u') e^{in(u-u')}-\int_0^{2\pi}du' \rho(u')e^{-in(u-u')})\\
    =\,&\frac{1}{2\pi}\int_0^{2\pi}du'\rho(u')\cot(\frac{u-u'}{2})
\end{align}

\bibliographystyle{jhep}
\bibliography{bibliography}

\end{document}